%% file: aa.tex
\documentclass{aa}

\usepackage{natbib}
\usepackage{graphicx}
\usepackage{txfonts}
\begin{document}

   \title{Deuterium Fractionation in the Ophiuchus Molecular Cloud\thanks{Based on observations carried out with the IRAM 30m Telescope. IRAM is supported by INSU/CNRS (France), MPG (Germany) and IGN (Spain).}}


   \author{A. Punanova
          \inst{1},
          P. Caselli\inst{1}, 
          A. Pon\inst{2},
          A. Belloche\inst{3}
          \and
          Ph. Andr\'e\inst{4}
          }

   \institute{Max-Planck-Institut f\"ur extraterrestrische Physik, Giessenbachstrasse 1, 85748 Garching, Germany\\
              \email{punanova@mpe.mpg.de}
         \and
             Department of Physics and Astronomy, The University of Western Ontario, London, ON, N6A 3K7, Canada
         \and
             Max-Planck-Institut f\"ur Radioastronomie, Auf dem Hügel 69, D-53121 Bonn, Germany
        \and
             IRFU/SAp CEA/DSM, Laboratoire AIM CNRS --- Universit\'e Paris Diderot, 91191 Gif-sur-Yvette, France
             }

   \date{Received October 19, 2015; accepted December 09, 2015}

 
  \abstract
   {In cold ($T < 25$\,K) and dense ($n_{H} > 10^4$\,cm$^{-3}$) interstellar clouds, molecules like CO are significantly frozen onto dust grain surfaces. Deuterium fractionation is known to be very efficient in these conditions as CO limits the abundance of H$_3^+$, the starting point of deuterium chemistry. In particular, ${\rm N_2D^+}$ is an excellent tracer of dense and cold gas in star forming regions. 
    }
   {We measure the deuterium fraction, R$_{\rm D}$, and the CO-depletion factor, $f_d$, toward a number of starless and protostellar cores in the L1688 region of the Ophiuchus molecular cloud complex and search for variations based upon environmental differences across L1688. The kinematic properties of the dense gas traced by the N$_2$H$^+$ and N$_2$D$^+$ (1--0) lines are also discussed.}
   {R$_{\rm D}$ has been measured  via observations of the J = 1--0 transition of N$_2$H$^+$ and N$_2$D$^+$ toward 33 dense cores in different regions of L1688. $f_d$ estimates have been done using C$^{17}$O(1--0) and 850~$\mu$m dust continuum emission from the SCUBA survey. All line observations were carried out with the IRAM 30~meter antenna.}
   {The dense cores show large ($\simeq$2--40\%) deuterium fractions, with significant variations between the sub-regions of L1688. The CO-depletion factor also varies from one region to another (between $\simeq$1 and 7). Two different correlations are found between deuterium fraction and CO-depletion factor: cores in regions A, B2 and I show increasing R$_{\rm D}$ with increasing $f_d$, similar to previous studies of deuterium fraction in pre-stellar cores;  cores in regions B1, B1B2, C, E, F and H show a steeper R$_{\rm D}$--$f_d$ correlation, with large deuterium fractions occurring in fairly quiescent gas with relatively low CO freeze-out factors. These are probably recently formed, centrally concentrated starless cores which have not yet started the contraction phase toward protostellar formation. We also find that the deuterium fraction is affected by the amount of turbulence, dust temperature and distance from heating sources in all regions of L1688, although no clear trend is found.}
   {The deuterium fraction and amount of CO freeze-out are sensitive to environmental conditions and their variations across L1688 show that regions of the same molecular cloud experience different dynamical, thermal and chemical histories, with consequences for the current star formation efficiency and the characteristics of future stellar systems. The large pressures present in L1688 may induce the formation of small dense starless cores, unresolved with our beam, where the R$_{\rm D}$--$f_d$ relation appears to deviate from that expected from chemical models. We predict that high angular resolution observations will reconcile observations with theory.}


   \maketitle
%

\section{Introduction}

The first stages of the star formation process are dense starless and self-gravitating cores, i.e. the so-called pre-stellar cores \citep{Ward-Thompson1999,Crapsi2005}. Pre-stellar cores in nearby star forming regions are typically cold ($\sim$10~K), dense (10$^4$--10$^7$~cm$^{-3}$) and quiescent \citep[thermal pressure dominates over turbulent motions; e.g.][]{Benson1989,Fuller1992,Lada2008,Keto2008}. 

Chemical differentiation takes place in pre-stellar cores \citep[see e.g.][for reviews]{Bergin2007,DiFrancesco2007,Caselli2011}. While CO is the second most abundant molecule in the interstellar medium, it tends to freeze onto dust grains in the dense, cold conditions at the centres of pre-stellar cores \citep[e.g.][]{Caselli1999,Bacmann2002}. The level of CO depletion is usually measured as $f_d$ = $X_{ref}({\rm CO})/N({\rm CO})\cdot N({\rm H_2})$, where $X_{ref}({\rm CO})$ is the reference value of the CO fractional abundance, typically between 1 and 2 $\times$ 10$^{-4}$ \citep[e.g.][]{Frerking1982,Lacy1994}. The typical value of the CO-depletion factor in pre-stellar cores is 5--20 \citep{Crapsi2005,Christie2012}.

In such cold and dense gas, deuterated species are preferentially formed \citep[e.g.][]{Caselli-Ceccarelli2012}. H$_2$D$^+$ is responsible for the enhancement of the deuterium fraction in most molecular species and is formed by the deuteron-proton exchange reaction H$^+_3$+HD$\rightleftharpoons$H$_2$D$^+$+H$_2$+230~K \citep{Millar1989}. The deuteron-proton exchange reaction is exothermic and does not proceed from right to left at temperatures lower than 30~K and if most of the H$_2$ molecules are in para form \citep[e.g.][]{Pagani1992}. H$_2$D$^+$ then reacts with other species to form deuterated ions via H$_2$D$^+$+A$\rightarrow$AD$^+$+H$_2$, where A can be any of CO, N$_2$ and other neutral species \citep{Herbst1973,Dalgarno1984}. When CO and other abundant neutral species, which destroy H$_3^+$ and H$_2$D$^+$, are severely frozen onto dust grain surfaces, the deuterium fraction becomes significant. For example, the deuterium fraction in pre-stellar cores is 5--50\%, while the elemental abundance of deuterium is $\sim$1.5$\times 10^{-5}$ with respect to hydrogen atoms within 1~kpc of the Sun \citep{Linsky2006,Caselli2011}. 

In particular, the deuterium fraction of N$_2$H$^+$ has been used to identify the earliest phases of star-formation, as the N$_2$H$^+$ deuterium fraction peaks at the pre-stellar phase and toward the youngest protostars \citep{Crapsi2005,Emprechtinger2009,Friesen2013,Fontani2014}. The deuterium fraction in N$_2$H$^+$ is usually given as R$_{\rm D}$ = $N({\rm N_2D^+})/N({\rm N_2H^+})$, where $N(i)$ is the column density of species $i$.

L1688 is a nearby, 120~pc distant \citep{Lombardi2008} low-mass star forming region within the Ophiuchus Molecular Cloud Complex. The multiple star forming regions within L1688 contain more than 60 dense cores and 50 young stellar objects in different evolutionary stages \citep{Motte1998,Andre2007,Simpson2008,Pattle2015}. L1688 is divided into 10 regions (A--I; see Fig.~\ref{Oph}) with different environmental properties. For instance, while the gas temperature is relatively constant within each region, it varies significantly from one region to another \citep[$\simeq$10--17~K;][]{Friesen2009}.

The deuterium fraction across the entirety of L1688 has not been systematically studied yet. For only a few regions has the deuterium fraction been measured, for example the B2 region has an average R$_{\rm D}\sim$3\% \citep{Friesen2010b}. 
CO-depletion across the whole of Ophiuchus has been found to be relatively low compared to the other Gould Belt star-forming regions, with an average value less than 10 \citep{Gurney2008,Christie2012}.

In this paper, we present observations of N$_2$D$^+$(1--0), N$_2$D$^+$(2--1), N$_2$H$^+$(1--0), C$^{17}$O(1--0) and C$^{17}$O(2--1) towards 40 cores to measure the deuterium fraction and CO depletion factor across the entire L1688 region. In Section 2, details regarding the observations are presented. Section 3 describes the results of hyperfine structure fitting as well as deuterium fraction and CO-depletion calculations. In Section 4, we discuss the results and their relation to possible environmental effects. The conclusions are given in Section 5. 


\section{Observations and data reduction}
\begin{figure*}\centering
\includegraphics[height=12cm,keepaspectratio]{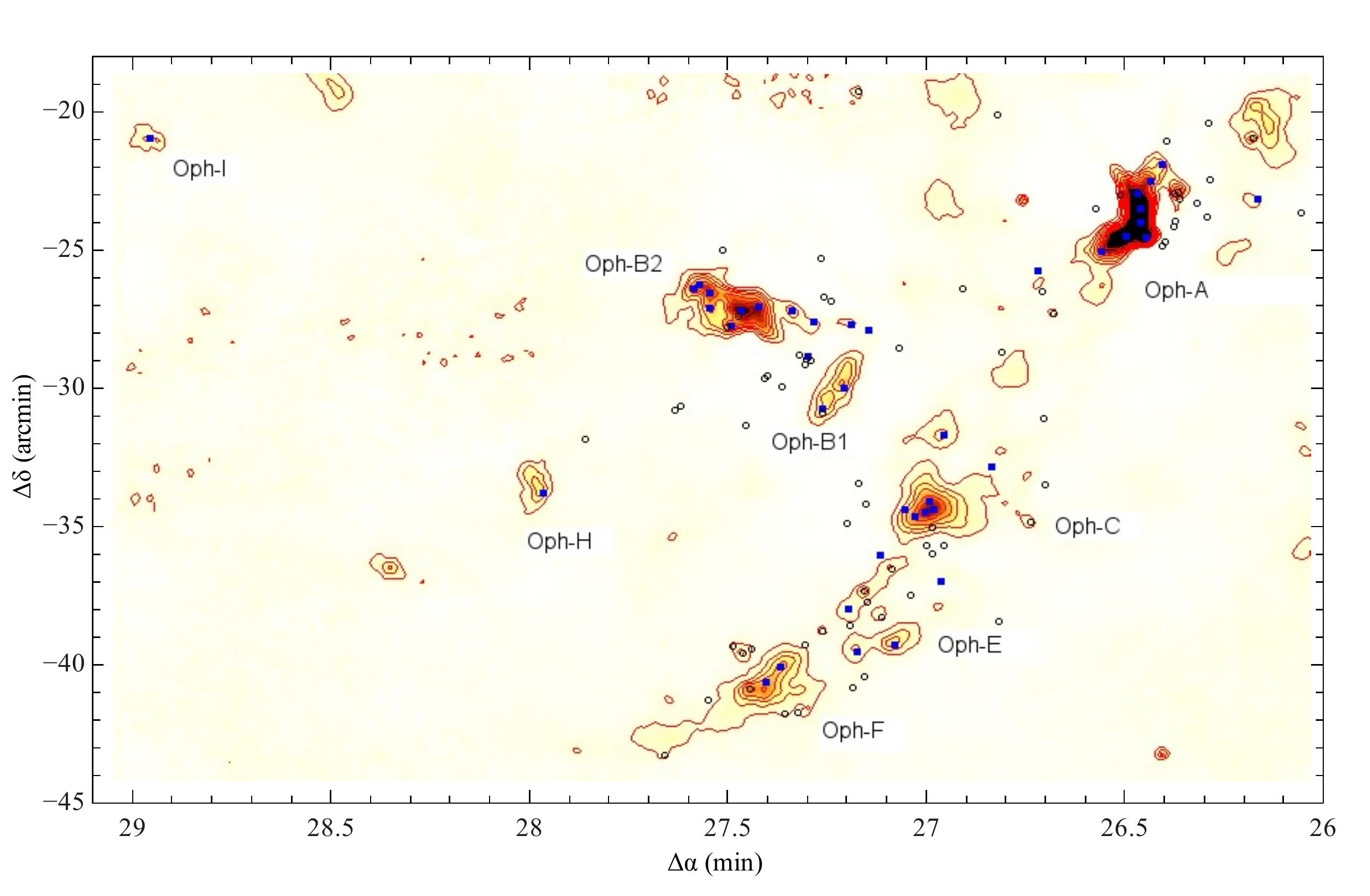}\centering
\caption{850~$\mu$m continuum emission of the L1688 region mapped by the Submillimeter Common-User Bolometer Array \citep[SCUBA,][]{DiFrancesco2008}, the beam size is 22.$^{\prime\prime}$9. Contour levels go from 0.2 Jy~beam$^{-1}$ in steps of 0.2~Jy~beam$^{-1}$ (3$\sigma$). The dense cores studied here are marked by filled blue squares and young stellar objects by open circles. The (0,0) offsets correspond to the J2000 equatorial position $\alpha=16^{\rm h}00^{\rm m}00^{\rm s}$, $\delta=-24^{\circ}00^{\prime}00^{\prime\prime}$. The fits file used to produce this map is available at http://www.cadc-ccda.hia-iha.nrc-cnrc.gc.ca/data/pub/JCMTSL/scuba\_F\_850umemi, file name  scuba\_F\_353d1\_16d8\_850um.emi.fits.}\label{Oph}
\end{figure*}

Figure~\ref{Oph} shows the L1688 region mapped in 850~$\mu$m dust continuum emission~\citep{DiFrancesco2008}. 40 dense cores, revealed by \citet{Motte1998} with 1.3~mm dust emission mapping, were selected for observation with the IRAM 30~meter telescope and are shown with filled blue squares. The names and positions of the cores are given in Table~\ref{lines}. Figure~\ref{Oph} also shows the positions of young stellar objects (YSOs) embedded in the cloud as open circles \citep{Motte1998,Simpson2008,Dunham2015}. The molecular line observations were performed with the IRAM 30~meter telescope in June 1998, July 2000 and December 2004. The following transitions were observed: N$_2$D$^+$(1--0), N$_2$D$^+$(2--1), N$_2$D$^+$(3--2), N$_2$H$^+$(1--0), N$_2$H$^+$(3--2), C$^{17}$O(1--0) and C$^{17}$O(2--1). The observations were obtained with the AB receiver and the VESPA backend. Typical system temperatures for the (1--0) transition observations were 100--200~K for N$_2$H$^+$--N$_2$D$^+$ and 200--360~K for the C$^{17}$O line. The spectral resolution for the N$_2$H$^+$, N$_2$D$^+$ and C$^{17}$O (1--0) lines varied from 6.5 to 40~kHz and the angular resolutions were 22, 32.1 and 26.6 arc seconds for C$^{17}$O(1--0), N$_2$D$^+$(1--0) and N$_2$H$^+$(1--0), respectively (see Table~\ref{observations}). The spectra were taken using the position switching (datasets 051-00 and 188-97) and frequency switching (dataset 066-04, with a frequency throw of 7.8~MHz) modes. In Table~\ref{observations}, the dates of the observing runs are given and each run denoted with a dataset number.  

The data reduction was performed with the CLASS package\footnote{Continuum and Line Analysis Single-Dish Software http://www.iram.fr/IRAMFR/GILDAS}. For each source, there were several spectra of the same line. These spectra have been adjusted to have the same central frequency and summed together to improve the sensitivity. The integration time for different lines and objects varies from 4 to 30~minutes. The intensity scale was converted to the main-beam temperature scale according to the beam efficiency values given in Table~\ref{observations}. 

The N$_2$D$^+$(1--0), N$_2$D$^+$(2--1), N$_2$H$^+$(1--0), C$^{17}$O(1--0) and C$^{17}$O(2--1) lines have hyperfine splitting with 15, 40, 15, 3 and 9 components respectively. 
As such, the spectra were analysed using the standard CLASS hyperfine structure (hfs) fitting method. The routine computes line profiles, with the assumptions of Gaussian velocity distribution and equal excitation temperatures for all hyperfine components. The rest frequencies of the main components, the velocity offsets and the relative intensities of the hyperfine components of the lines were taken from \citet{Frerking1981}, \citet{Pagani2009} and Dore,~L. (private communication). The N$_2$D$^+$(3--2) and N$_2$H$^+$(3--2) spectra have very poor baselines and reconstruction of the signal is not possible, so these data are not considered hereafter in the paper. 

All spectra were initially fit assuming one velocity component. The hfs fitting routine returns both the rms of the baseline and the region with the spectral line. In case the rms of the spectral line region was greater than the rms of the baseline by a factor of 1.5, we redid the fit with an additional velocity component. This was repeated until the two rms agreed within a factor of 1.5. The largest number of velocity components needed was three. 

\begin{table*}
\caption{Observation parameters.}\label{observations}\centering
\begin{tabular}{cccccccccc}
\hline
\hline
Species & Frequency$^a$ & F$_{eff}$ & B$_{eff}^f$ & HPBW & ${\rm \Delta v_{res}}^g$ & rms in T$_{mb}$ & T$_{sys}$	& Dates& Dataset$^h$\\
        & (GHz)    &         &           &($^{\prime\prime}$) &(km~s$^{-1}$)& (K)& (K)&   & \\
\hline
N$_2$H$^+$(1--0) & 93.1737637$^b$ & 0.95 & 0.76 & 26.5 & 0.021 & 0.121 & 155--176 & 16.08.2004 & 066-04\\
N$_2$H$^+$(1--0) & 93.1737637 & 0.92 & 0.78 & 26.5 & 0.063 & 0.089 & 114--205 & 11--16.07.2000 & 051-00\\
N$_2$H$^+$(1--0) & 93.1737637 & 0.92 & 0.73 & 26.5 & 0.063 & 0.174 & 173--216 & 26--28.06.1998 & 188-97\\
N$_2$H$^+$(3--2) & 279.511832$^b$ & 0.87 & 0.46 & 8.8 & 0.021 &    & 652--2739 & 12--13.08.2004 & 066-04 \\
N$_2$D$^+$(1--0) & 77.1096162$^b$ & 0.95 & 0.76 & 32.1 & 0.025 & 0.103 & 167--216 & 12--17.08.2004 & 066-04\\
N$_2$D$^+$(2--1) & 154.2171805$^c$ & 0.94 & 0.64 & 16.3 & 0.013 & 0.225 & 203--627 & 12--15.08.2004 & 066-04 \\
N$_2$D$^+$(3--2) & 231.3219119$^c$ & 0.90 & 0.52 & 10.8 & 0.025 &       & 244--710 & 16.08.2004 & 066-04 \\
C$^{17}$O(1--0)  & 112.358988$^d$ & 0.95 & 0.78 & 22.0 & 0.017 & 0.218 & 243--320 & 16.08.2004 & 066-04\\
C$^{17}$O(1--0)  & 112.358988 & 0.92 & 0.78 & 22.0 & 0.052 & 0.143 & 184--289 & 12.07.2000 & 051-00\\
C$^{17}$O(1--0)  & 112.358988 & 0.92 & 0.73 & 22.0 & 0.052 & 0.180 & 265--359 & 29--30.06.1998 & 188-97\\
C$^{17}$O(2--1)  & 224.714370$^e$ & 0.85 & 0.53 & 11.0 & 0.532 & 0.626 & 629--1366 & 13-16.07.2000 & 051-00\\
C$^{17}$O(2--1)  & 224.714370 & 0.90 & 0.42 & 11.0 & 0.052 & 0.977 & 995-1800 & 29--30.06.1998 & 188-97\\
\hline
\end{tabular}
\\
\begin{flushleft}
$^a$ Frequency of the main hyperfine component;
$^b$ from \citet{Pagani2009};
$^c$ from \citet{Pagani2009} and Dore,~L., private communication;
$^d$ from \citet{Frerking1981};
$^e$ from SPLATALOGUE database http://www.cv.nrao.edu/php/splat/;
$^f$ B$_{eff}$ values are available at the 30~m antenna efficiencies web-page https://www.iram.fr/IRAMFR/ARN/aug05/node6.html;
$^g {\rm \Delta v_{res}} $ is the velocity resolution;
$^h$ dataset name is the ID of the corresponding IRAM 30~m project.
\end{flushleft}
\end{table*}

\section{Results}

\subsection{Spectra}

\begin{figure*}\centering
\includegraphics[height=23cm,keepaspectratio,width=\hsize]{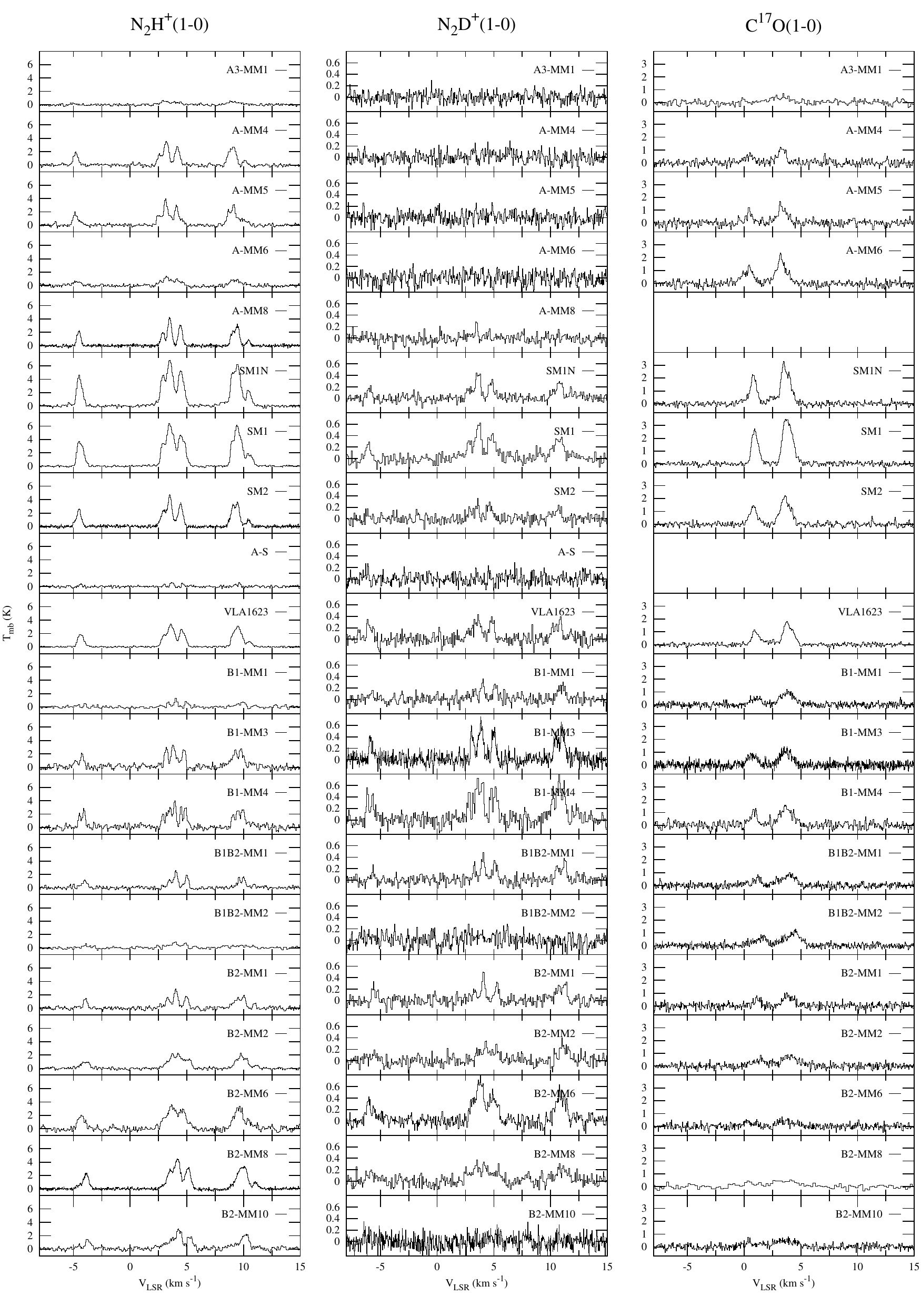}
\caption{N$_2$H$^+$(1-0), N$_2$D$^+$(1-0) and C$^{17}$O(1--0) spectra toward the observed dense cores, labeled in each panel.}\label{spectra-1}
\end{figure*}
\addtocounter{figure}{-1} 
\begin{figure*}\centering 
\includegraphics[height=23cm,keepaspectratio,width=\hsize]{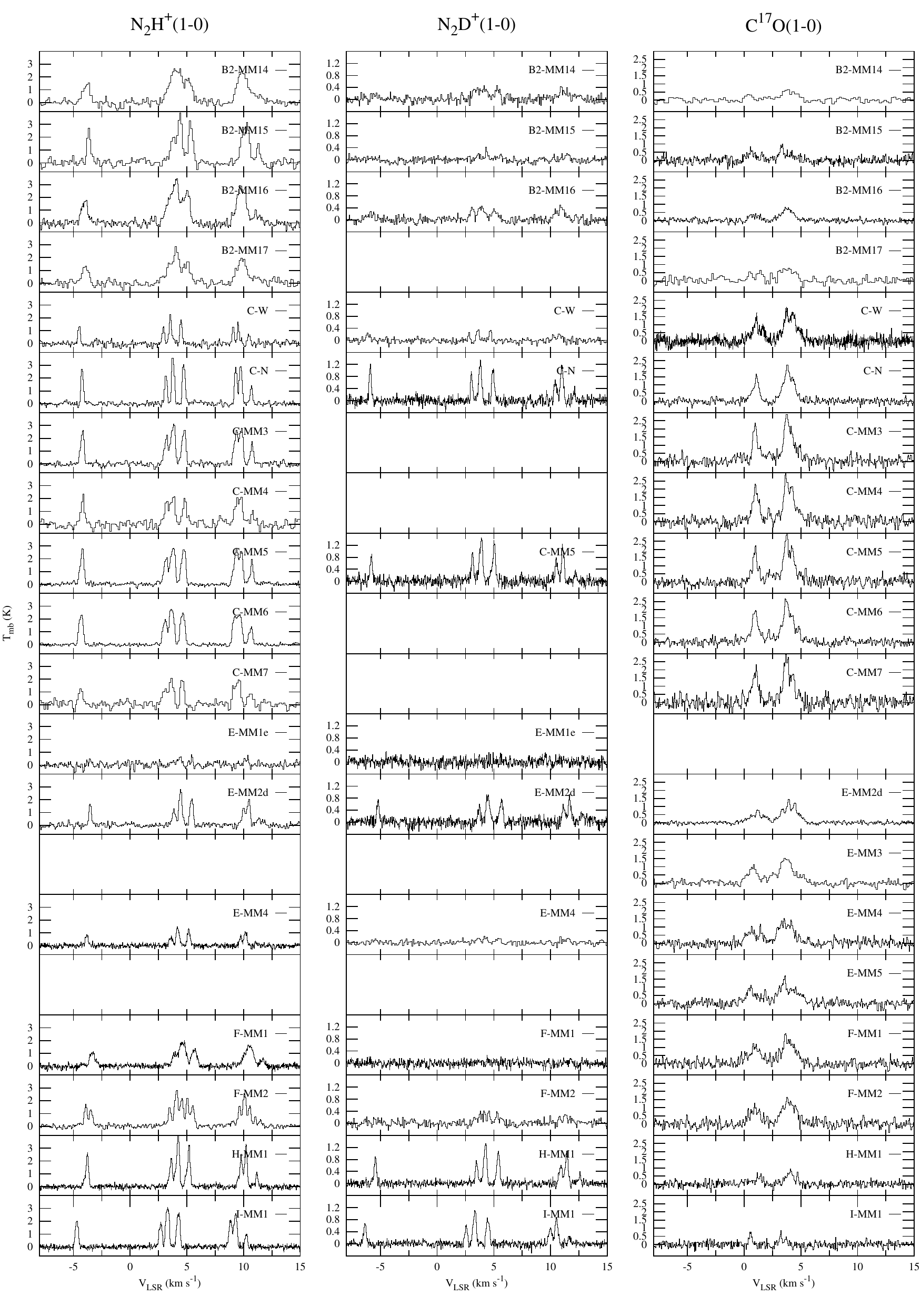}
\caption{N$_2$H$^+$(1-0), N$_2$D$^+$(1-0) and C$^{17}$O(1--0) spectra toward the observed dense cores, labeled in each panel (continuation).}\label{spectra-2}
\end{figure*}

The spectra of the N$_2$D$^+$(1--0), N$_2$H$^+$(1--0) and C$^{17}$O(1--0) lines are shown in Fig.~\ref{spectra-1}. 
The N$_2$H$^+$(1--0), C$^{17}$O(1--0) and C$^{17}$O(2--1) emission was detected toward all 40 observed cores. N$_2$D$^+$(1--0) emission was detected toward 23 out of 33  observed cores and N$_2$D$^+$(2--1) emission was detected toward 25 out of 32 observed cores, with the A-MM4 core only having (2--1) detection. As the aim of the study is the measurement of deuterium fractions and their comparison to CO-depletion factors, we focus on the (1--0) transitions for the remainder of the paper, since they have the most similar beam sizes and excitation conditions. Toward five cores (B1-MM3, B1-MM4, B2-MM2, B2-MM8 and F-MM2), the N$_2$H$^+$(1--0) line shows two velocity components. The C$^{17}$O(1--0) line toward all of the objects in regions C (except C-Ne and C-MM3) and E and one in A (SM1N) shows two or three velocity components. The ${\rm N_2D^+(1-0)}$ line shows two velocity components toward one core, B1-MM4. The results of the hfs fits are given in Tables~\ref{N2H-plus-results}--\ref{CO21-results}. 

\begin{figure*}\centering
\begin{minipage}{0.49\linewidth}
\includegraphics[height=7.5cm,keepaspectratio]{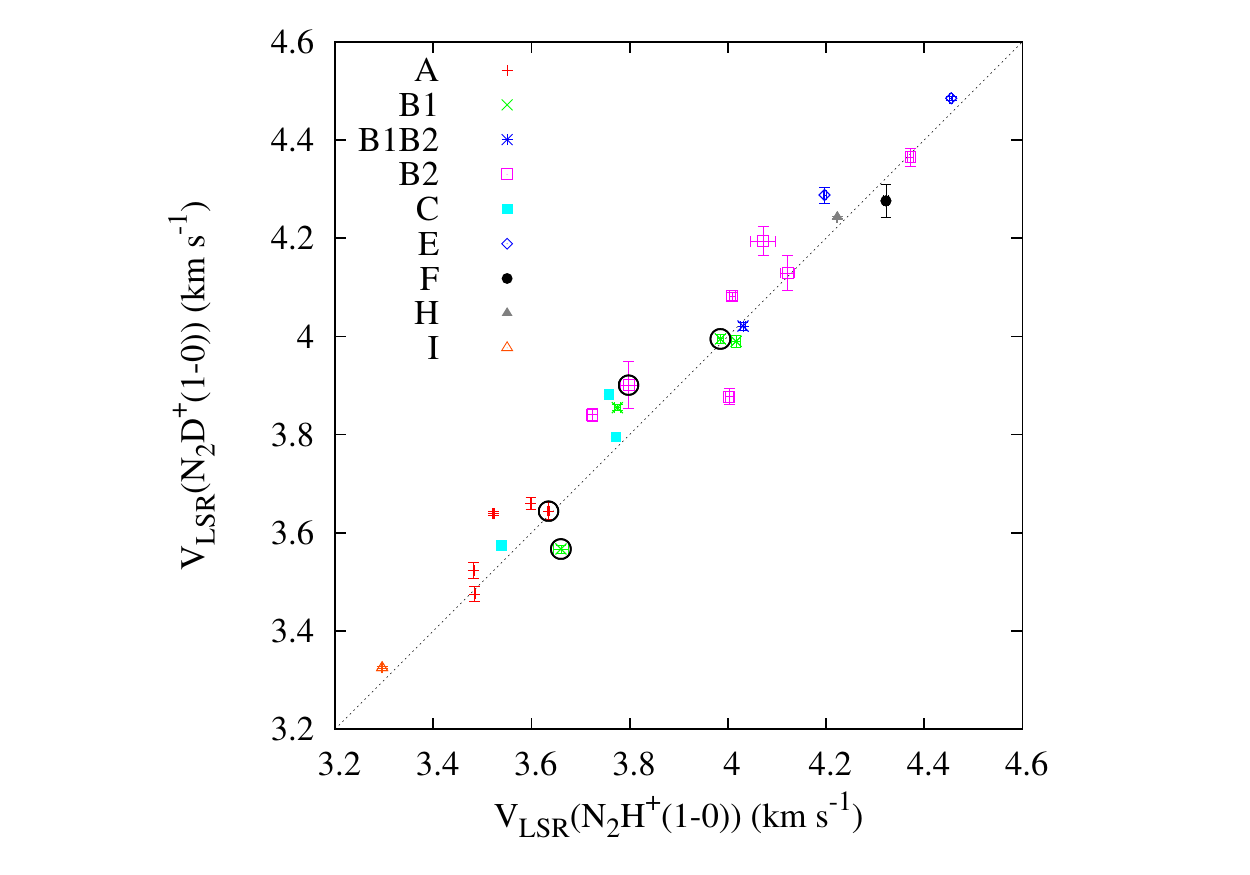}
\end{minipage}
\hfill
\begin{minipage}{0.49\linewidth}
\includegraphics[height=7.5cm,keepaspectratio]{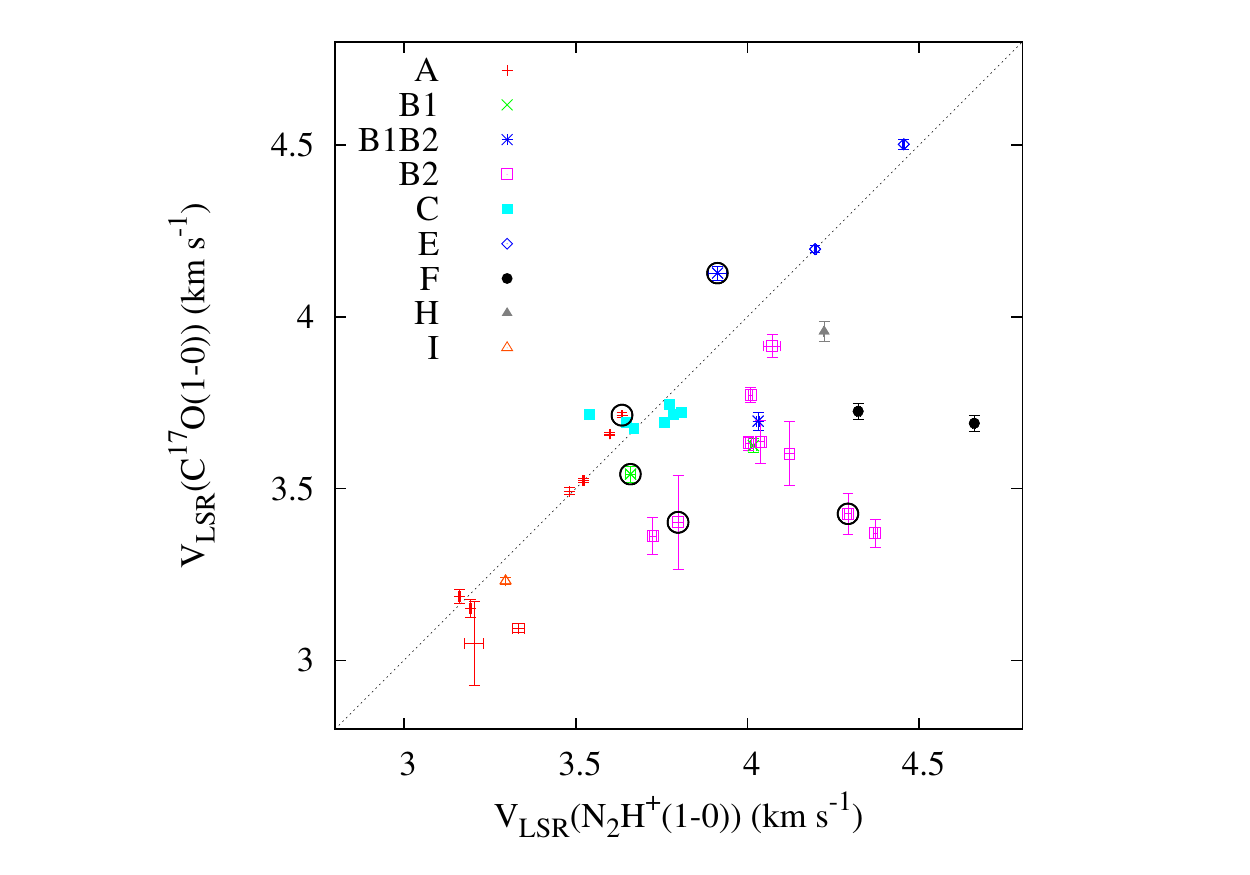}
\end{minipage}
\caption{V$_{\rm LSR}$ of N$_2$D$^+$(1--0) on the left panel and C$^{17}$O(1--0) on the right panel as a function of the V$_{\rm LSR}$ of the N$_2$H$^+$(1--0) line. Black dotted lines are the lines of equal V$_{\rm LSR}$. The protostellar cores are marked with black open circles.}\label{VLSR-N2D-N2H}. 
\end{figure*}

\begin{figure}\centering
\includegraphics[height=6cm,keepaspectratio,width=\hsize]{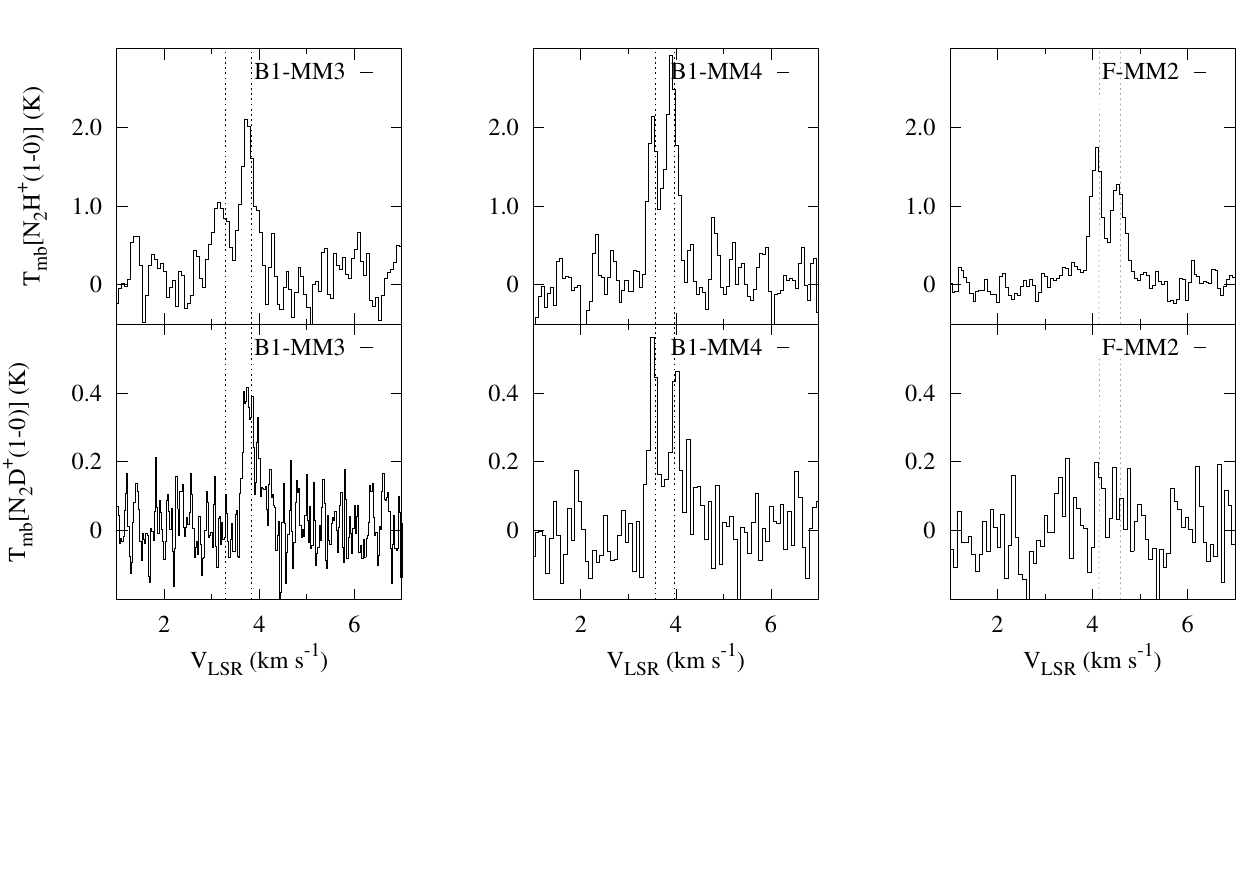}
\caption{Isolated hyperfine components of N$_2$H$^+$ and N$_2$D$^+$ (1--0) toward cores where two velocity components are found. The spectra are centred at the frequency of the isolated component. The dotted lines show the centroid velocities of the N$_2$H$^+$(1--0) components.}\label{sin-com-1}
\end{figure}

The centroid velocities, V$_{\rm LSR}$, are determined from the hfs fitting and vary across L1688 from 3.3 to 4.6~km~s$^{-1}$ with a velocity generally increasing from region A to F. Figure~\ref{VLSR-N2D-N2H} shows the centroid velocities for the N$_2$H$^+$(1--0), N$_2$D$^+$(1--0) and C$^{17}$O(1--0) lines. If two velocity components are detected in N$_2$H$^+$(1--0) and N$_2$D$^+$(1--0) (e.g. toward B1-MM4), both components are plotted in the left panel of Fig.~\ref{VLSR-N2D-N2H}. If only N$_2$H$^+$(1--0) shows two components we instead plot an average of the two components if the single N$_2$D$^+$(1--0) line component appears to be a blend of multiple component (cores B2-MM2, B2-MM8 and F-MM2), and we plot just the velocity of the closest N$_2$H$^+$(1--0) component if the N$_2$D$^+$(1--0) component does not appear to be blended, with the second N$_2$D$^+$(1--0) component presumably not being detected above the noise level (core B1-MM3). The three cases are illustrated in Fig.~\ref{sin-com-1}. However, only a small fraction of the points (4/25) shown in Fig.~\ref{VLSR-N2D-N2H} are for locations with different numbers of components detected in the two different tracers. For the right panel of Fig.~\ref{VLSR-N2D-N2H}, we plot each N$_2$H$^+$(1--0) component against the closest C$^{17}$O(1--0) component.  

For 80\% of the cores, the V$_{\rm LSR}$ of the N$_2$H$^+$(1--0) and N$_2$D$^+$(1--0) lines are within 0.05~km~s$^{-1}$ of each other. The largest centroid velocity difference is only 0.21~km~s$^{-1}$. The C$^{17}$O(1--0) and N$_2$H$^+$(1--0) V$_{\rm LSR}$ can differ significantly with discrepancies up to 1~km~s$^{-1}$. 51\% of the cores have velocities that differ by over 0.1~km~s$^{-1}$. This suggests that N$_2$H$^+$(1--0) and N$_2$D$^+$(1--0) trace roughly the same gas while C$^{17}$O(1--0) likely traces more extended gas, as expected considering the widespread distribution of CO in molecular clouds and its freeze-out in dense cold regions.

\begin{figure*}\centering
\begin{minipage}{0.49\linewidth}
\includegraphics[height=7.5cm,keepaspectratio]{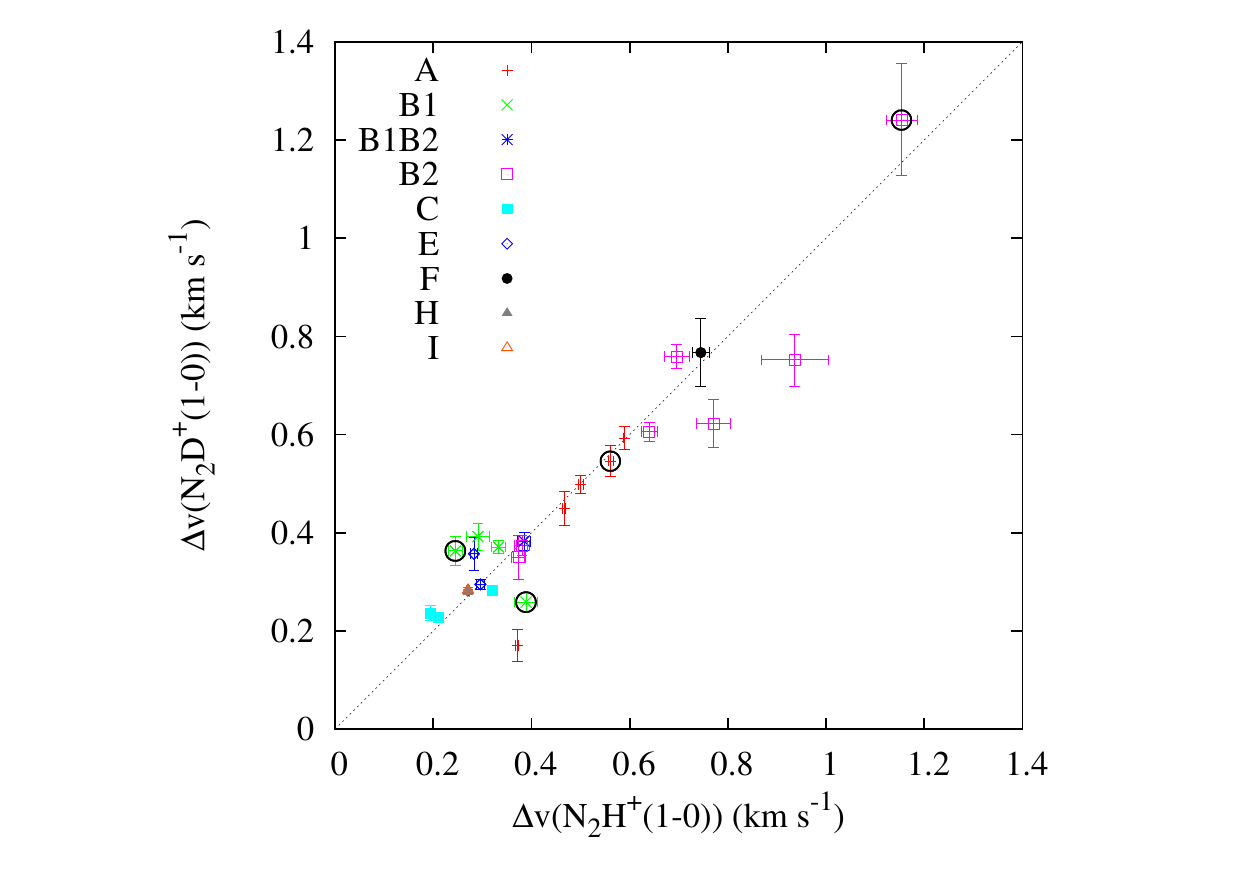}
\end{minipage}
\hfill
\begin{minipage}{0.49\linewidth}
\includegraphics[height=7.5cm,keepaspectratio]{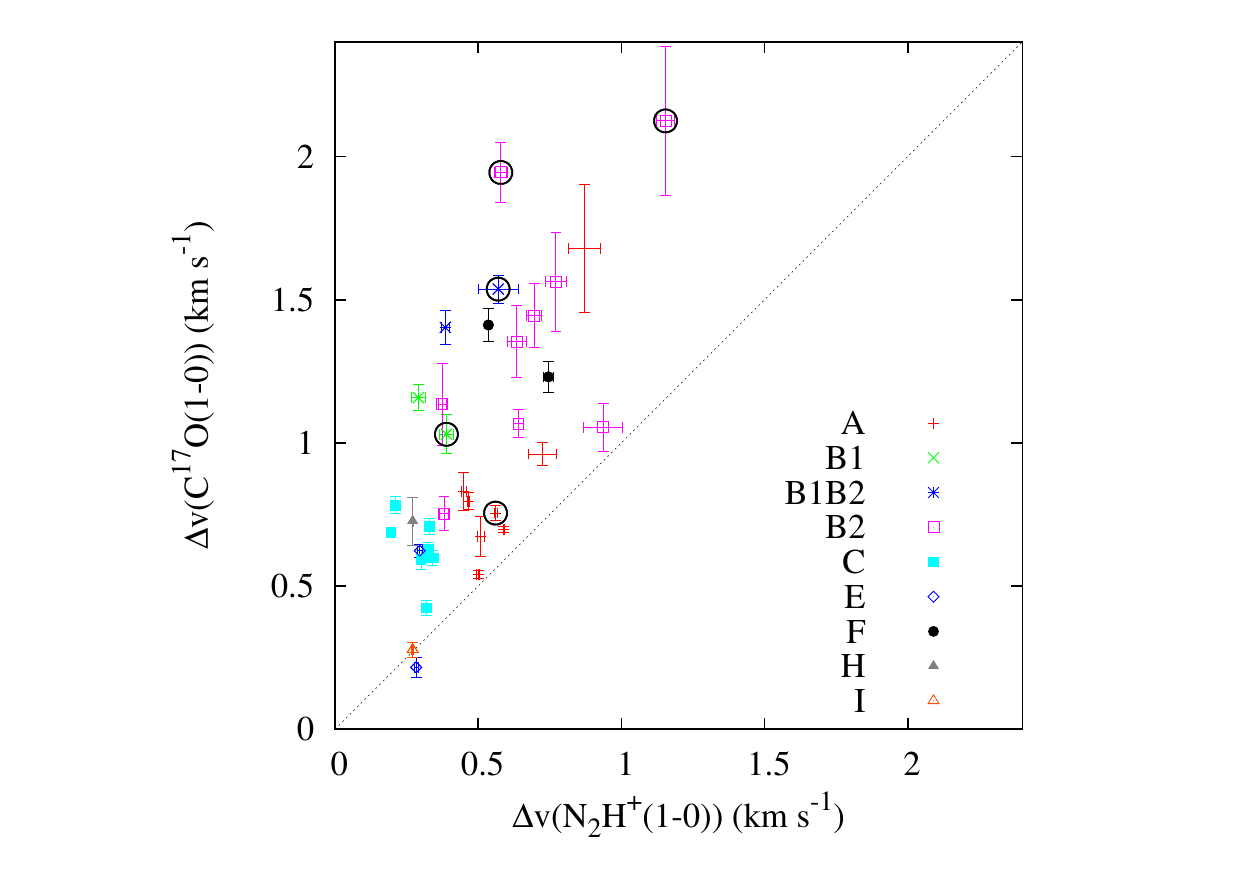}
\end{minipage}
\caption{Line widths, $\Delta$v, of N$_2$D$^+$(1--0) on the left panel and C$^{17}$O(1--0) on the right panel in comparison with $\Delta$v of N$_2$H$^+$(1--0). Black dotted lines are the lines of equal $\Delta$v. Different colors show different parts of the cloud. The protostellar cores are marked with black open circles.}\label{fwhm}
\end{figure*}

The line widths (full width at half maximum, FWHM, hereinafter $\Delta{\rm v}$) of N$_2$H$^+$, N$_2$D$^+$ and C$^{17}$O (1--0) are shown on Fig.~\ref{fwhm}. As before, when N$_2$H$^+$ and N$_2$D$^+$ show two velocity components, we plot them separately (B1-MM4). In the case where two velocity components are present in N$_2$H$^+$ and only one is seen in N$_2$D$^+$, we either take the $\Delta$v of N$_2$H$^+$ of the component with the closest V$_{\rm LSR}$, if it appears that one component of N$_2$D$^+$ is missing due to the noise level (B1-MM3), or we take $\Delta$v=$\Delta$v$_1$/2+$\Delta$v$_2$/2+|V$_{\rm LSR 1}$-V$_{\rm LSR 2}$| if the N$_2$D$^+$ line appears to be a blend of two components (B2-MM2, B2-MM8, F-MM2). The three cases are shown in Fig.~\ref{sin-com-1}. For the comparison of the N$_2$H$^+$ and C$^{17}$O line widths, we took the $\Delta$v of the component having the closest V$_{\rm LSR}$. 

The N$_2$H$^+$ and N$_2$D$^+$ (1--0) line widths range from 0.2 to 0.7~km~s$^{-1}$, except for the blended N$_2$D$^+$ line at B2-MM8 (1.2~km~s$^{-1}$, see left panel of Fig.~\ref{fwhm}). While the line widths of the N$_2$H$^+$ and N$_2$D$^+$ (1--0) lines are similar for most of the cores, 84\% of the cores are within 0.1~km~s$^{-1}$ of each other, the C$^{17}$O(1--0) line widths are overall larger than those of N$_2$H$^+$(1--0). The median width difference is 0.38~km~s$^{-1}$ and the median line width ratio is 1.9. Similar to what was found with the line centroids, this suggests that the N$_2$H$^+$ and N$_2$D$^+$ trace the same gas, while C$^{17}$O traces different, in particular more turbulent, gas. 

\subsection{Non-thermal motions}

\begin{figure}
\includegraphics[height=6cm,keepaspectratio]{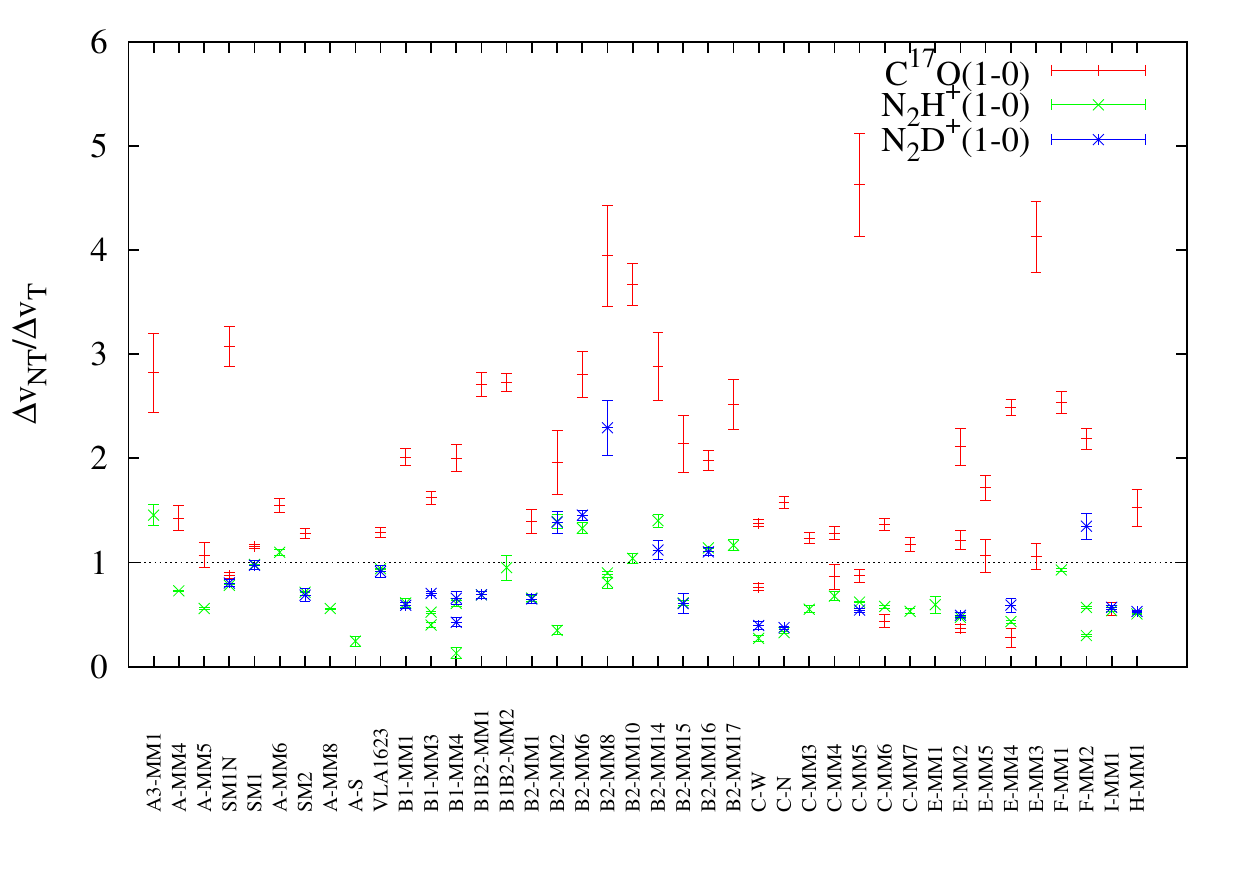}
\caption{Ratio of non-thermal components to thermal components of the N$_2$D$^+$(1--0), N$_2$H$^+$(1--0) and C$^{17}$O(1--0) lines. The dashed line shows the ratio equal to 1.}\label{ntmotions}
\end{figure}

Figure~\ref{ntmotions} presents the ratio of non-thermal components $\Delta {\rm v_{NT}}$ of the N$_2$D$^+$(1-0), N$_2$H$^+$(1-0) and C$^{17}$O(1-0) lines and thermal line widths of a mean particle, $\Delta{\rm v_T}$. The non-thermal components are derived from the observed line widths $\Delta{\rm v}_{obs}$ via: 
\begin{equation}
\Delta{\rm v}^2_{\rm NT}=\Delta{\rm v}^2_{obs}-8\ln(2)\frac{kT_k}{m_{obs}},
\end{equation}
where $k$ is Boltzmann's constant, $T_k$ is the kinetic temperature, and $m_{obs}$ is the mass of the observed molecule \citep{Myers1991}. 
To measure the non-thermal component, we use the kinetic temperature determined by \citet{Friesen2009} from ammonia observations. For those cores which were not observed in \citet{Friesen2009}, we use dust temperatures determined by \citet{Pattle2015}, assuming that the dust and gas temperatures are equivalent \citep[assumption valid at volume densities above 10$^4$~cm$^{-3}$;][]{Goldsmith2001}. For most of the cores where both the gas and dust temperatures have been measured, the two values are indeed similar, with the only exception being the B2 region, where the dust temperature is a few degrees lower than the gas temperature. This may be due to the effect of protostellar feedback, where shocks produced by outflows entraining the dense gas can heat the gas but not the dust \citep[e.g.][]{Draine1980}.  
For the I-MM1 core neither dust nor gas temperatures have ever been estimated, so we adopt 11~K, the same as H-MM1, as these two cores have similar characteristics, both being relatively isolated and far away from the main source of irradiation and heating (see Section\,\ref{section_heating}). The kinetic temperatures for all cores are given in Table~\ref{N-RD-fD-results}.

For typical temperatures of 10--20~K across L1688, the thermal line widths, $\Delta$v$_T$, for a mean particle with mass 2.33\,amu are 0.44--0.63~km~s$^{-1}$.
The majority of C$^{17}$O(1--0) lines are supersonic (78\%) while most of N$_2$H$^+$(1--0) and N$_2$D$^+$(1--0) lines are subsonic (80\% and 75\%). The non-thermal to thermal line width ratio can be as high as 1.5 for N$_2$H$^+$(1--0), 2.5 for N$_2$D$^+$(1--0), and as high as 5 for C$^{17}$O(1--0).
The most turbulent region is Oph-B2 and the most quiescent regions are Oph-B1, C, E, H and I. Oph-A contains turbulent as well as relatively quiescent cores. For most cores, N$_2$D$^+$(1--0) and N$_2$H$^+$(1--0) have similar non-thermal components and are narrower than the C$^{17}$O(1--0) line. 

\subsection{Column densities and deuterium fractions}\label{sec:N-RD}

\begin{figure}\centering
\includegraphics[height=6cm,keepaspectratio]{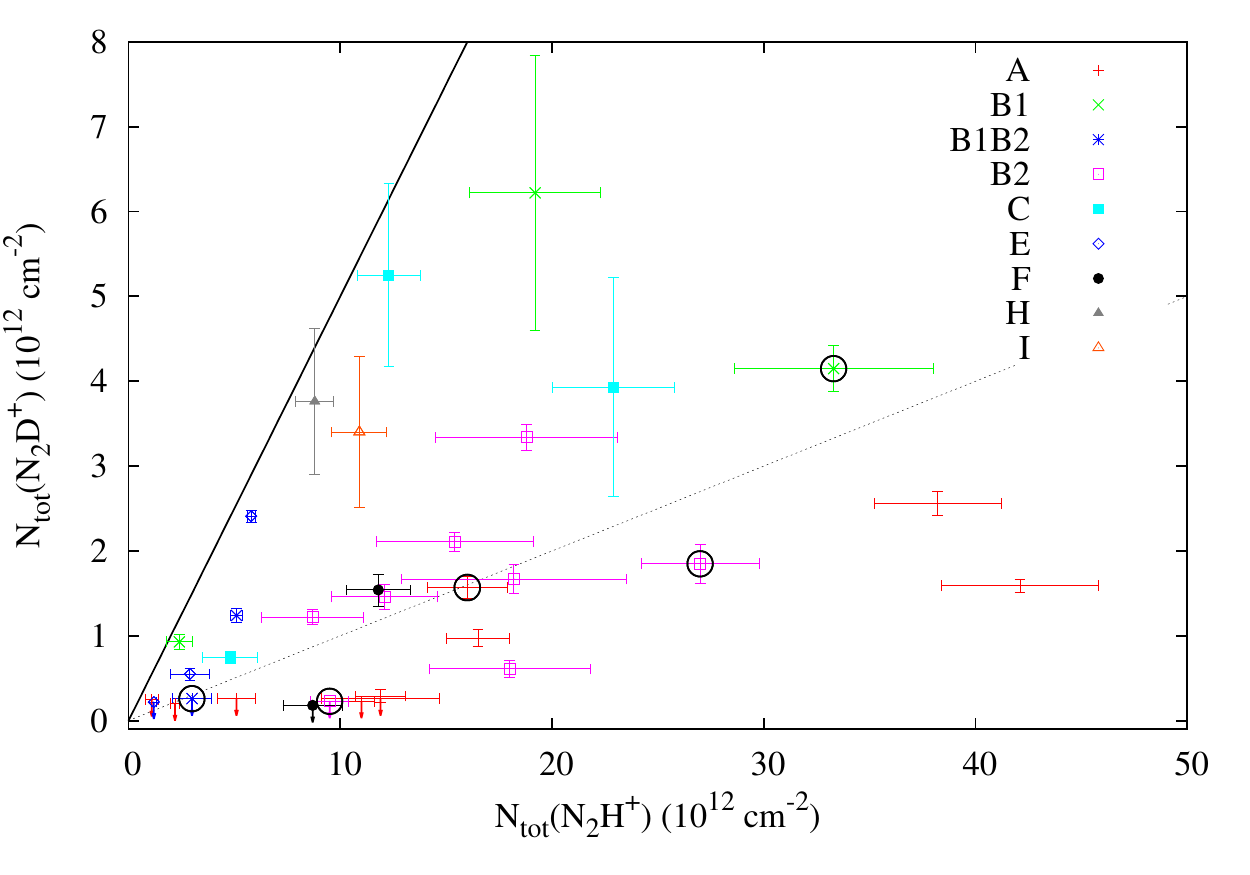}
\caption{Column densities of N$_2$D$^+$ and N$_2$H$^+$ with lines of constant deuterium fraction (${\rm R_D}$). The black line shows ${\rm R_D}$=0.5 and the dotted line shows ${\rm R_D}$=0.1. The protostellar cores are depicted with black open circles.}\label{Ntot}
\end{figure}

The hfs fits provide values needed to measure the excitation temperature ($T_{ex}$) and its error. These values are: the total optical depth, i.e. the sum of the optical depths of the various hyperfine components ($\tau$), the quantity labelled $T_{ant}\times\tau$ (see below), the full width at half maximum of the line (FWHM, $\Delta$v) and the centroid velocity relative to the local standard of rest (V$_{\rm LSR}$). In case of optically thick lines, $T_{ant}\times\tau$ is the total optical depth times the difference between the Rayleigh-Jeans equivalent excitation and background temperatures, while for optically thin lines it is the main beam temperature ($T_{mb}$). The $T_{ex}$ can be calculated as

\begin{equation}\label{Tex}
T_{ex}=\frac{h\nu}{k}\left[\ln\left(\frac{h\nu/k}{(T_{ant}\times\tau)/\tau+J_{\nu}(T_{bg})}+1\right)\right]^{-1},
\end{equation}
where $h$ is the Planck constant, $k$ is the Boltzmann constant, $\nu$ is the frequency of the observed transition, $T_{bg}$ is the cosmic background temperature ($2.7~$K), $J_{\nu}(T_{bg})$ is the equivalent Rayleigh-Jeans background temperature, and $J_{\nu}(T)$ is the function

\begin{equation}\label{Jnu}
J_{\nu}(T)=\frac{h\nu/k}{\exp(h\nu/kT)-1}.
\end{equation}

The calculated excitation temperature depends on the value of $\tau$. In the case of weak lines or low $S/N$, $\tau$ can not be determined properly and the error of $\tau$ ($\Delta\tau$) will be high. In all cases where $\tau/\Delta\tau\le 3$, we consider the lines to be optically thin and fix $\tau = 0.1$ (the minimum opacity value) in CLASS. In this case of optically thin conditions, for ${\rm N_2H^+(1-0)}$, the excitation temperature value is assumed to be the average $T_{ex}$ found for optically thick ${\rm N_2H^+(1-0)}$ lines, while for ${\rm N_2D^+(1-0)}$ we adopt the (measured or assumed) ${\rm N_2H^+(1-0)}$ excitation temperature toward the same dense core.

For optically thick transitions, the column density ($N_{tot}$) is given by:
\begin{equation}\label{Nthick}
N_{tot}=\frac{8\pi^{3/2}\Delta {\rm v}}{2\sqrt{\ln 2}\lambda^3A_{ul}}\frac{g_l}{g_u}\frac{\tau}{1-\exp(-h\nu/kT_{ex})}\frac{Q_{rot}}{g_l\exp(-E_l/kT_{ex})},
\end{equation}
where $\lambda$ is the wavelength of the observed transition, $A_{ul}$ is the Einstein coefficient of the $u\rightarrow l$ transition, $g_l$ and $g_u$ are the statistical weights of the lower and upper levels, $Q_{rot}$ is the partition function and $E_l$ is the energy of the lower level \citep{Caselli2002-b}. For linear rotors, $g_l$ and $g_u$ are determined by $g_J=2J+1$, where $J$ is the rotational quantum number. The partition function of linear molecules (such as N$_2$H$^+$ and CO) is given by
\begin{equation}
Q_{rot}=\sum^\infty_{J=0}(2J+1)\exp(-E_J/kT),
\end{equation}
where $E_J=J(J+1)hB$, and $B$ is the rotational constant. For rotational transitions with hyperfine structure, $\tau$ refers to the total optical depth (given by the sum of the peak optical depths of all the hyperfine components) and $\Delta$v to the intrinsic line width. The error on $N_{tot}$ is given by propagating the errors on $\Delta$v, $\tau$ and $T_{ex}$ in equation~\ref{Nthick}.\par
For optically thin lines
\begin{equation}
\begin{split}
N_{tot}=\frac{8\pi W}{\lambda^3A_{ul}}\frac{g_l}{g_u}\frac{1}{J_{\nu}(T_{ex})-J_{\nu}(T_{bg})}\frac{1}{1-\exp(-h\nu/kT_{ex})} \times \\
\times \frac{Q_{rot}}{g_l\exp(-E_l/kT_{ex})},
\end{split}
\end{equation}
where $W$ is the integrated intensity of the line:
\begin{equation}
W=\frac{\sqrt{\pi}\Delta v T_{mb}}{2\sqrt{\ln 2}},
\end{equation}
for a Gaussian line \citep{Caselli2002-b}.

In case of non-detection of N$_2$D$^+$, upper limits on the N$_2$D$^+$ column density have been derived based on the $3\sigma$ uncertainty ($3\,\sigma_W$) of the integrated intensity, with:
\begin{equation}
\sigma_W=rms \times \sqrt{N_{ch}} \times {\rm \Delta v_{res}} ,
\end{equation} 
where $N_{ch}$ is the mean number of channels covering the velocity range of all the detected lines and ${\rm \Delta v_{res}} $ is the velocity resolution.

The column densities of N$_2$H$^+$, N$_2$D$^+$ and C$^{17}$O are given in tables~\ref{N2H-plus-results}--\ref{CO21-results}. The N$_2$D$^+$ column densities derived from the (1--0) transition in most cores are larger than those derived from the (2--1) transition on average by only 10\%. C$^{17}$O column densities calculated with (1--0) lines in most cores are smaller than the ones calculated with (2--1) lines (see section~\ref{sec:fd} for details). Figure~\ref{Ntot} shows the column densities of ${\rm N_2H^+}$ and ${\rm N_2D^+}$. Where multiple components are detected, we plot the sum of the column densities.

\begin{figure}\centering
\includegraphics[height=6cm,keepaspectratio]{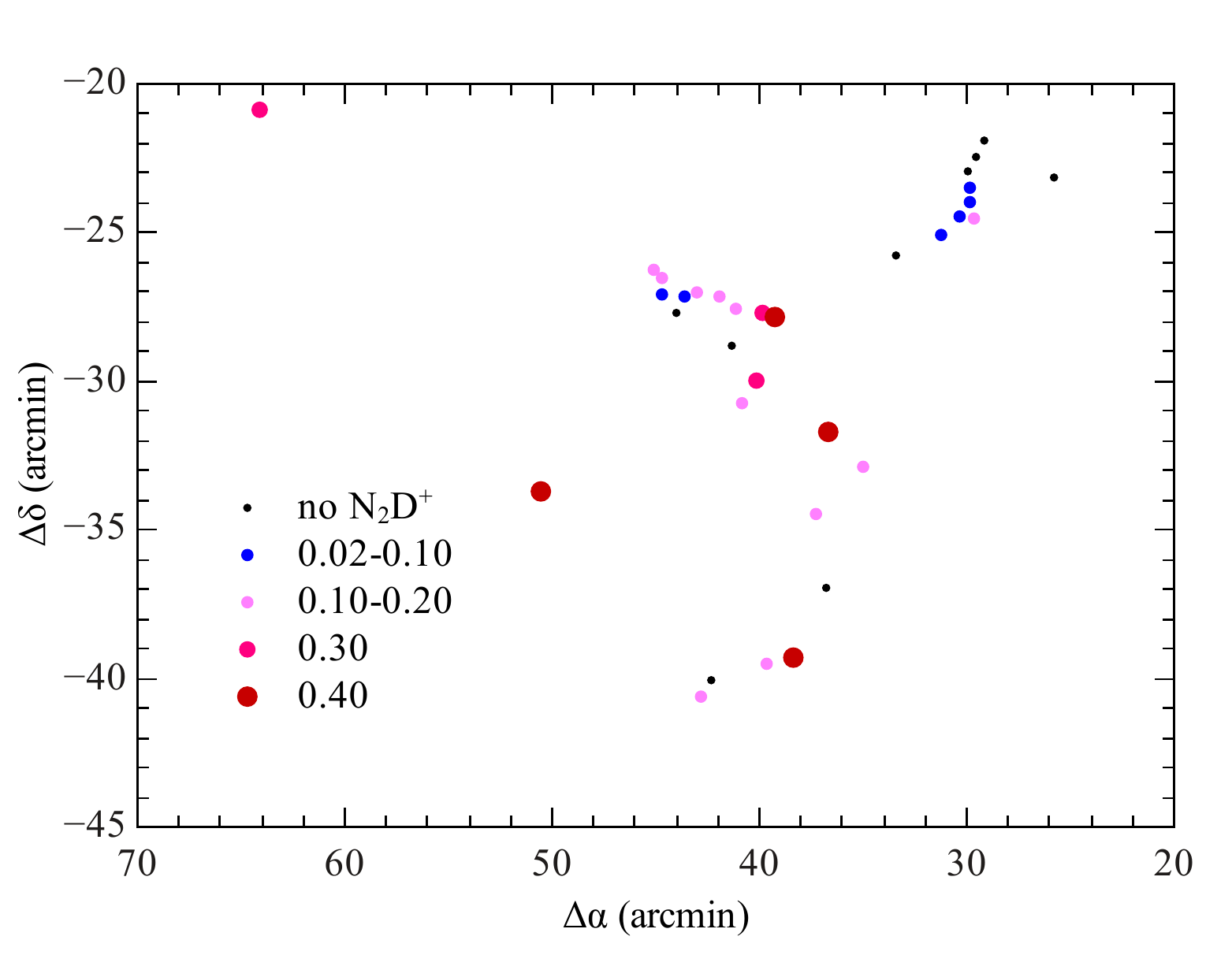}
\caption{Deuterium fraction across L1688. The black points show the cores with no N$_2$D$^+$(1--0) detection. The coordinate offsets correspond to the J2000 equatorial position $\alpha=16^{\rm h}24^{\rm m}16^{\rm s}$, $\delta=-24^{\circ}00^{\prime}00^{\prime\prime}$.}\label{RD-map}
\end{figure}

The deuterium fraction is defined as the ratio of column densities, R$_{\rm D}={N_{tot}{\rm (N_2D^+)}/N_{tot}{\rm (N_2H^+)}}$, and it has been measured for all cores where the N$_2$D$^+$(1--0) line is detected. To calculate the deuterium fraction for cores where two velocity components are detected, we take the sum of the column densities derived from the two components. L1688 overall exhibits high levels of deuterium fractions with a large spread of values between different cores (${\rm R_D}=2$--76~\%; see Fig.~\ref{RD-map}). Such a high level of deuteration (over 20~\%) was previously found toward other dense cores in different star forming regions \citep[e.g.][]{Crapsi2005,Pagani2007,Emprechtinger2009,Fontani2011,Miettinen2012,Friesen2013,Fontani2014}. The deuterium fraction across the B2 region was previously studied by \citet{Friesen2010b}, who found slightly lower deuterium fractions (1--10\%) than in the present work (3--18\%). The column densities of N$_2$D$^+$ (0.5--6$\times$10$^{11}$~cm$^{-2}$) and N$_2$H$^+$ (4--10$\times$10$^{12}$~cm$^{-2}$) they obtain are also smaller than those found in this work (6--33$\times$10$^{11}$ and 9--27$\times$10$^{12}$~cm$^{-2}$). This  difference could be due to different transitions used to calculate column densities, while using similar excitation temperatures derived from the N$_2$H$^+$(1--0) line. \citet{Friesen2010b} used N$_2$D$^+$(3--2) and N$_2$H$^+$(4--3) lines with 11$^{\prime\prime}$ and 13$^{\prime\prime}$ HBPW while we use the (1--0) transition for both species (32$^{\prime\prime}$ and 27$^{\prime\prime}$).  The factor of 2 difference in deuterium fraction could arise also from not coincident dense core coordinates. 

\subsection{CO-depletion factor}\label{sec:fd}

In cold, dense, quiescent gas, CO freezes out onto dust grains and the level of this depletion is commonly expressed as a CO-depletion factor, $f_d$, calculated as:

\begin{equation}\label{fD}
{f_d}=\frac{X_{ref}({\rm CO})}{X({\rm CO})},
\end{equation}
where $X_{ref}({\rm CO})$ is the reference abundance and $X({\rm CO})$ is the observed abundance.

The reference abundance of CO in the local ISM has been found to be between 1 and 2$\times 10^{-4}$ \citep{Wannier1980,Frerking1982,Lacy1994}. We use $X_{ref}({\rm C^{16}O})=2\times 10^{-4}$ \citep{Frerking1982},  $X({\rm C^{18}O})/X({\rm C^{17}O})=4.11$ \citep{Wouterloot2005} and $X({\rm C^{16}O})/X({\rm C^{18}O})=560$ \citep{Wilson1994} such that:

\begin{equation}\label{XCO}
X({\rm C^{16}O})=\frac{N_{tot}({\rm C^{17}O})\times 4.11\times 560}{N({\rm H_2})}.
\end{equation}

To calculate the $f_d$ in case of multiple velocity components, we consider the sum of the column densities of the individual components (as $N({\rm H_2})$ is derived from the millimetre dust continuum emission, which does not contain kinematic information).
Since millimetre dust continuum emission is generally optically thin, the molecular hydrogen column density can be derived from the continuum flux density:

\begin{equation}\label{NH2}
N({\rm H_2})=\frac{S_{\lambda}}{\Omega \, \mu_{\rm H_2} \, m_{\rm H} \, \kappa_{\lambda} \, B_{\lambda}(T_{dust})};
\end{equation}

\noindent
where $S_\lambda$ is the flux in a single beam, $\Omega$ is the main beam solid angle, $\mu_{\rm H_2}=2.8$ is the mean molecular weight per H$_2$ molecule \citep{Kauffmann2008}, $m_{\rm H}$ is the mass of atomic hydrogen, $\kappa_\lambda$ is the dust opacity per unit mass column density at a given wavelength \citep[$\kappa_{850 \mu m}$=0.01~cm$^2$~g$^{-1}$;][]{Johnstone2000} and $B_{\lambda}(T_{dust})$ is the Planck function for a dust temperature $T_{dust}$ \citep{Motte1998}. 

 The 850~$\mu$m dust continuum emission flux measurements from the SCUBA survey \citep[with the beam size of the convolved map of 22$^{\prime\prime}$.9,][]{DiFrancesco2008} have been used to calculate $N(\rm H_2)$, adopting the dust temperature from \citet{Pattle2015} when available and the gas temperature from   \citet{Friesen2009} in the other cases. For those cores not studied in the above mentioned papers, the dust temperature estimated by \citet{Motte1998} has been adopted. For Oph-I, where no dust or gas temperature has been measured, we assumed 11~K, the same as in Oph-H, as Oph-I has similar characteristics to Oph-H, as already mentioned. The dust temperatures for all cores are given in Table~\ref{N-RD-fD-results}.

The depletion factor of CO is generally quite low in L1688, ranging from 0.2 to 2 in the B1, B1B2, C, E, and F regions, and from to 0.7 to 7.3 in the A, B2, H and I regions. This result is consistent with previous large scale \citep[e.g.][]{Christie2012} and small scale \citep[e.g.][]{Bacmann2002,Gurney2008} studies. However, larger CO depletion factors are found for the A region compared to the work of \citet{Gurney2008}: 1 to 6 instead of 1.5 to 4.5. This small discrepancy could be due to slightly different pointings or from the use of different CO isotopologues and transitions. In particular, we note that our C$^{17}$O(2--1) observations tend to produce column densities larger than $N_{tot}$(C$^{17}$O(1--0)) by an average factor of 1.5 in Oph-A and 1.1 in B1, B1B2, B2, C, E and F regions (and thus depletion factors would be lower by 1.5 and 1.1), suggesting that temperature and density gradients along the line of sight may be present, slightly affecting the derived depletion factor depending on the CO transition used. To calculate column densities of ${\rm C^{17}O}$, LTE is assumed, with the kinetic temperature equal to the dust temperature. 

\citet{Christie2012} suggest that the low depletion factor of L1688 could be due to an unusual dust grain size distribution with a population of very large dust grains and very small spinning dust grains which reduces the surface area available for freeze-out. However, L1688 is a complex region, with active star formation and externally irradiated. Below, we discuss possible causes of the general low CO depletion factors and the significant variation in ${\rm R_D}$ and $f_d$ found across L1688. 

\section{Discussion}

\subsection{Deuterium fraction}

This work has found a large range of deuterium fractions across L1688, from a minimum of 2\% to a maximum of 43\%. Previous studies of deuterium fraction did not show such a big spread: for example, 5--25\% in Taurus \citep{Crapsi2005}, 1--10\% in Ophiuchus B \citep{Friesen2010b}, and 3--25\% in Perseus \citep{Emprechtinger2009,Friesen2013}.  The large values of deuterium fractions found toward some of the cores in L1688 in our more extensive survey may indicate the presence of centrally concentrated pre-stellar cores on the verge of star formation. The Ophiuchus Molecular Cloud is known to be denser on average than other nearby star forming regions \citep[e.g.][]{Lada2013}. The higher average densities, together with the generally higher molecular cloud temperatures \citep{Pattle2015,Liseau2015}, imply larger pressures which may accelerate the formation of denser cores and the rate of star formation \citep[e.g.][]{Kennicutt2012} compared to the other nearby molecular cloud complexes. It is interesting to note that these high R$_{\rm D}$ values are found toward E-MM2d, C-Ne, I-MM1, H-MM1, B1-MM1, B1-MM3, B1B2-MM1 which do not prominently appear in the 850\,$\mu$m map and they are all relatively isolated structures (I and H) or in between bright sub-millimetre clumps (E, C-N, B1, B1B2). They are probably recently formed cold and dense structures on the verge of star formation. 

\subsection{Deuterium fraction and CO depletion}\label{sec:RD-fd}

\begin{figure}\centering
\includegraphics[height=6cm,keepaspectratio]{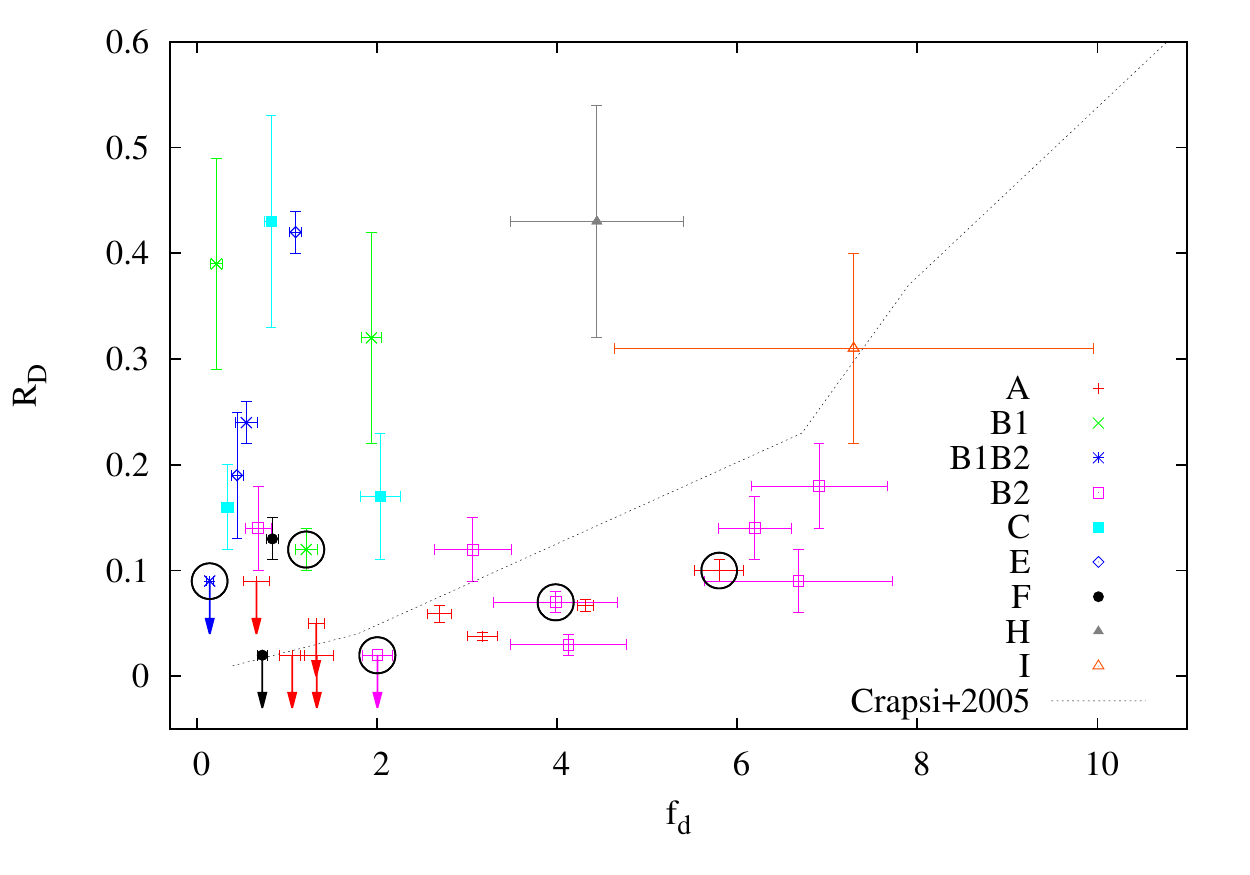}
\caption{Deuterium fraction as a function of CO-depletion factor. The protostellar cores are depicted with black open circles.}\label{RD-fD}
\end{figure}

\begin{figure*}\centering
\begin{minipage}{0.49\linewidth}
\includegraphics[height=6cm,keepaspectratio]{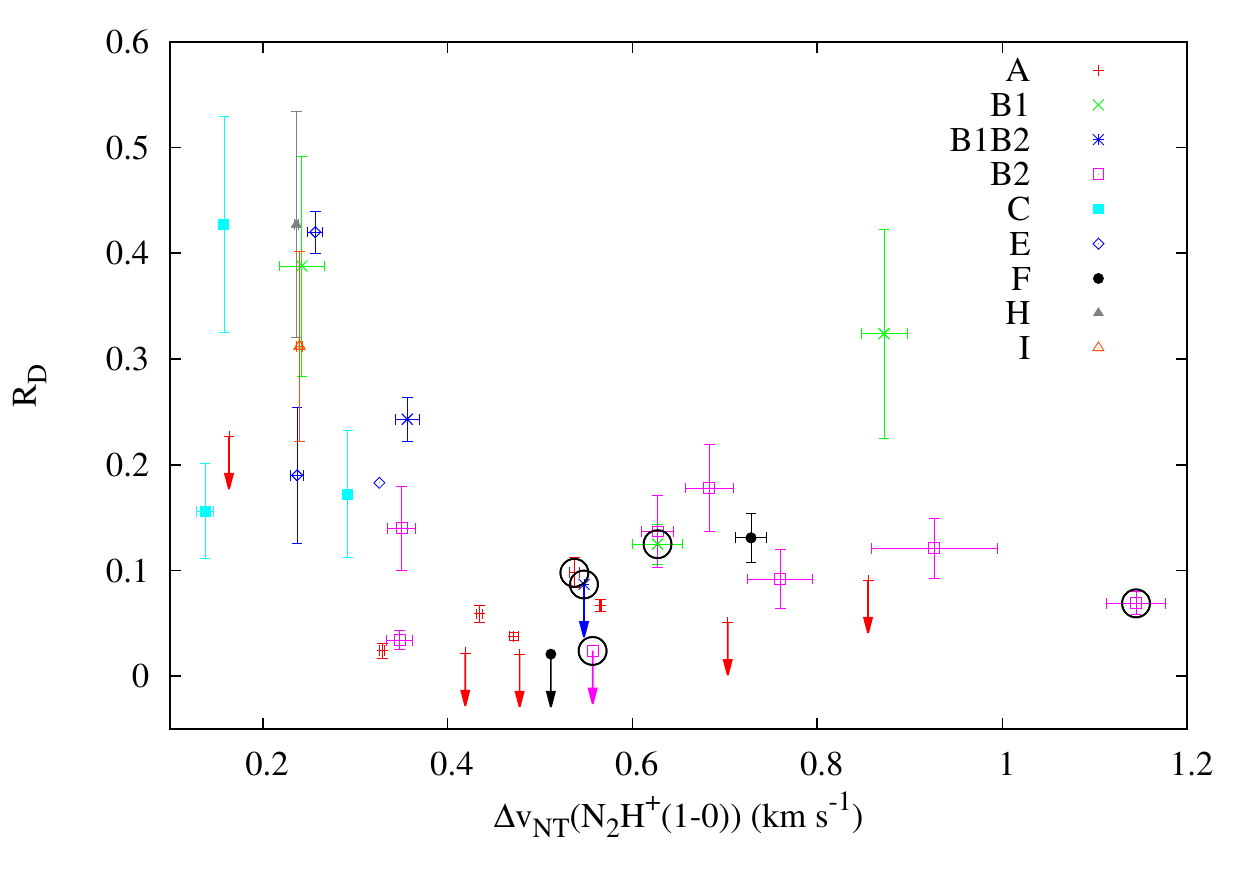}
\end{minipage}
\hfill
\begin{minipage}{0.49\linewidth}
\includegraphics[height=6cm,keepaspectratio]{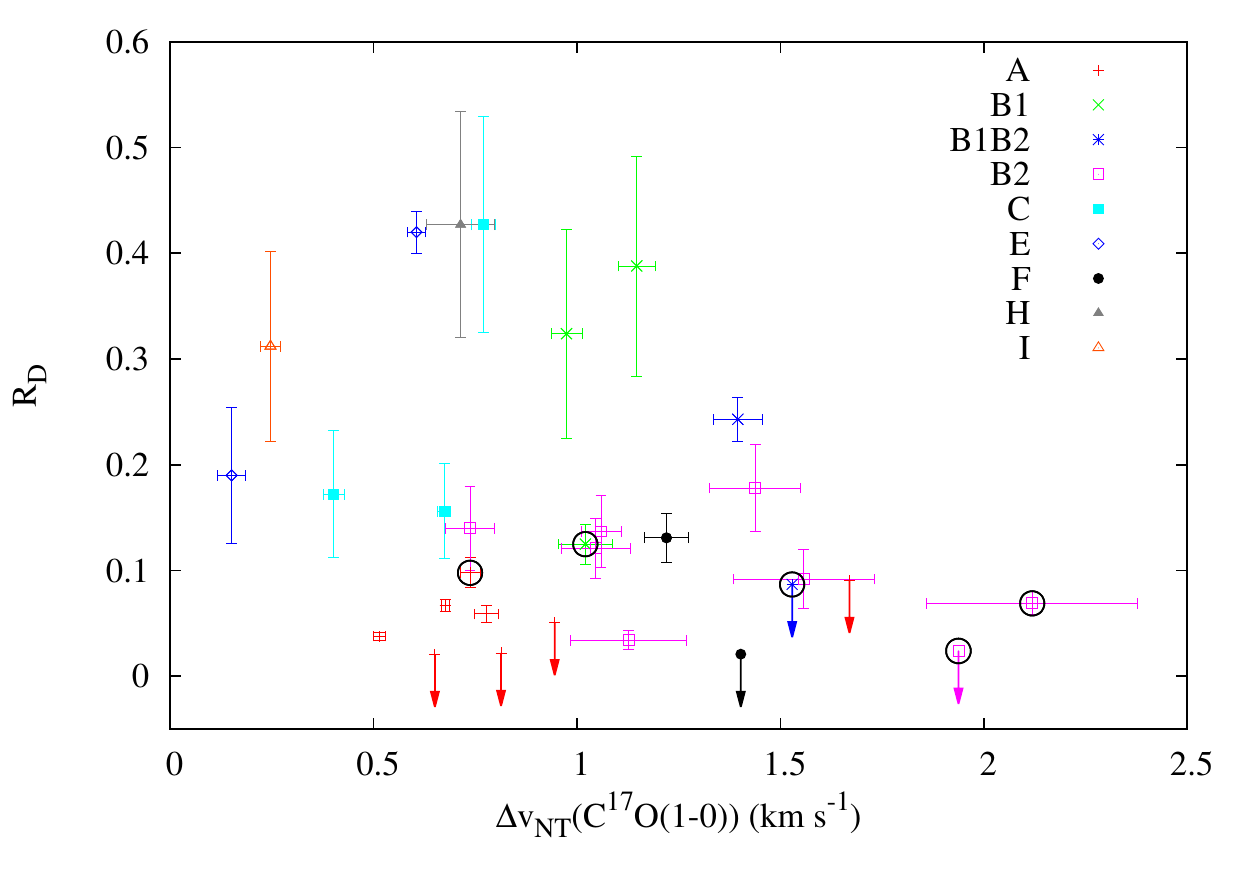}
\end{minipage}
\caption{Deuterium fraction as a function of the nonthermal component  of the N$_2$H$^+$(1--0) line width (left) and the C$^{17}$O(1--0) line width (right).  The protostellar cores are depicted with black open circles.}\label{RD-fwhm}
\end{figure*}

Deuterium fraction and CO-depletion factor are expected to correlate, because CO is one of the main destruction partners of H$_3^+$ and its deuterated forms \citep{Dalgarno1984}. This correlation has been presented in several theoretical works \citep{Crapsi2005,Caselli2008,Kong2015} and confirmed with observations \citep{Crapsi2005,Emprechtinger2009,Friesen2013}. Figure~\ref{RD-fD} shows the deuterium fraction as a function of CO-depletion factor in L1688. The dotted curve on the figure shows the prediction from simple modelling by \citet{Crapsi2005}, shifted to take into account the different reference CO abundance adopted here. The first thing to note is that $f_d$ values go below 1, which suggests that our adopted $X_{ref}$(CO) has been underestimated by a factor of a few (2--3, considering the lowest $f_d$ value in Table~\ref{N-RD-fD-results}); or the dust opacity, $\kappa$, which depends on the evolutionary stage and the properties of dust grains \citep[see][]{Henning1995} has been overestimated by a factor of 1.5--2, or the dust temperature has been overestimated by a few (1--3)~K. However, to allow comparison with recent literature work \citep[where $X_{ref}$(CO) = 2$\times$10$^{-4}$, see e.g.][]{Hernandez2011,Fontani2012}, and considering the factor of 2 uncertainty associated with $X_{ref}$(CO) \citep[see also][]{Miotello2014}, we did not modify  $X_{ref}$(CO), warning the reader that the calculated $f_d$ values may be underestimated by a factor of a few.  The important message here is to see if previously found trends are reproduced and/or if L1688 hosts dense cores with a larger variety of chemical/physical properties than found in other nearby star forming regions.  

From Fig.~\ref{RD-fD}, it is evident that two classes of cores are present in L1688. One group includes the A, B2 and I cores, which show a correlation between ${\rm R_D}$ and $f_d$ similar to that found by \citet{Crapsi2005}. The other group contains the B1, B1B2, C, E, F and H cores, which completely deviate from the \citet{Crapsi2005} correlation. This latter group of cores shows ${\rm R_D}=$12--43\% and ${f_d}=$0.2--4.4. From the parameter space exploration of \citet{Caselli2008} and \citet{Kong2015}, large values of R$_{\rm D}$ ($>$ 0.02) cannot be achieved in standard conditions if little CO freeze-out is present {\it in the same gas traced by N$_2$D$^+$}.  How to reconcile theory with observations?  One possibility is that the dense and cold regions responsible for the bright N$_2$D$^+$ lines have sizes smaller than the IRAM-30m beam at 3mm.  In this case, the CO-depleted zone would be too diluted to be clearly detected within the larger scale CO-emitting region. Indeed, 2 of 6 cores with ${\rm R_D}>20$\% and $f_d<4.4$ (B1-MM3 and B1B2-MM1) have estimated sizes 1300--1800~AU, less then 2640~AU corresponding to 22$^{\prime\prime}$ at 120~pc, one unresolved (B1-MM1) \citep{Motte1998}, one (H-MM1) has no size estimate \citep[outside of the mapped area in][]{Motte1998}. Higher angular resolution observations of dust continuum and molecular lines are needed to prove this point. 

\subsection{Deuterium fraction and non-thermal motions}

\begin{figure}\centering
\includegraphics[height=6cm,keepaspectratio]{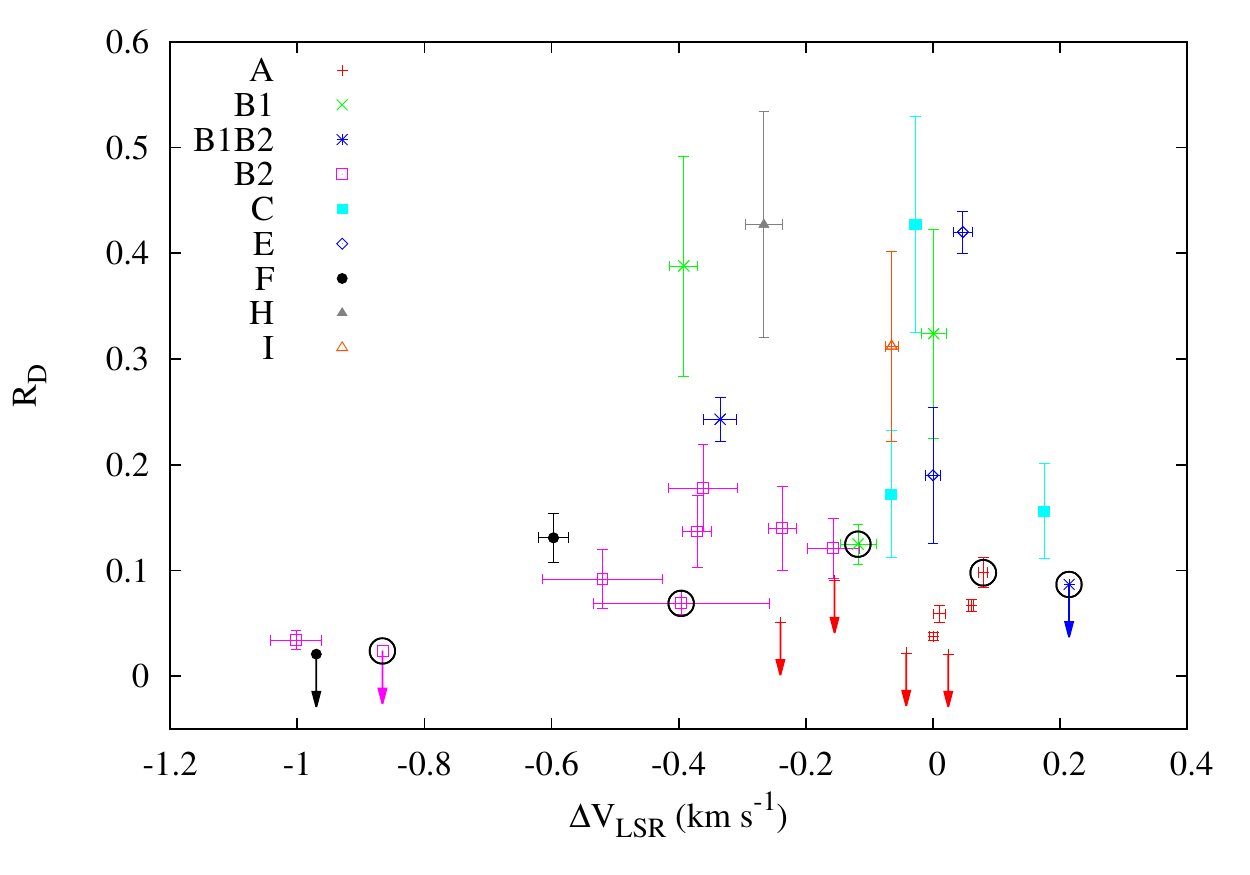}
\caption{Deuterium fraction as a function of the difference between the central velocities of the C$^{17}$O(1--0) and N$_2$H$^+$(1--0)  lines $\Delta{\rm V_{LSR}}$.  The protostellar cores are depicted with black open circles.}\label{del-VLSR}
\end{figure}

To understand further the characteristics of the highly deuterated, but CO-rich cores in Fig.~\ref{RD-fD}, we plot the deuterium fraction as a function of non-thermal line width of N$_2$H$^+$(1--0) and C$^{17}$O(1--0) in Fig.~\ref{RD-fwhm}. This figure shows that these cores preferentially occupy the left area of the panels, indicating that on average they have narrower C$^{17}$O(1--0) and especially N$_2$H$^+$(1--0) lines.  Thus, the highly deuterated cores are overall more quiescent than the rest of the sample, in agreement with their relatively isolated nature and maybe smaller size, as mentioned in the previous section.  Please also note that these are the cores with relatively small differences between C$^{17}$O(1--0) and N$_2$H$^+$(1--0) LSR velocities (see Fig.~\ref{del-VLSR}), again suggesting quiescent conditions. The rest of the sample displays broader line-widths, suggestive of faster internal motions (in case of gravitational contraction) or external stirring, e.g. due to proximity to active sites of star formation. Indeed, relatively large N$_2$H$^+$(1--0) line widths have been found by \citet{Crapsi2005} toward some of the most evolved starless cores in their sample (L1544 and L429; see their Fig.~6).  Line widths tend to increase toward the centre of L1544 \citep{Caselli2002-a} because of contraction motions.  The quiescent and highly deuterated cores found in L1688 may then represent an earlier evolutionary stage, compared to L1544 and other contracting pre-stellar cores, where the core has just started to become centrally concentrated but contraction has not started yet (or it has not affected scales large enough to be detected with the current single-dish observations). High angular resolution observations are needed to investigate this conclusion.   

\subsection{R$_{\rm D}$ and $f_d$ versus molecular hydrogen column density and temperature}\label{FandT}

\begin{figure*}\centering
\includegraphics[height=12cm,keepaspectratio]{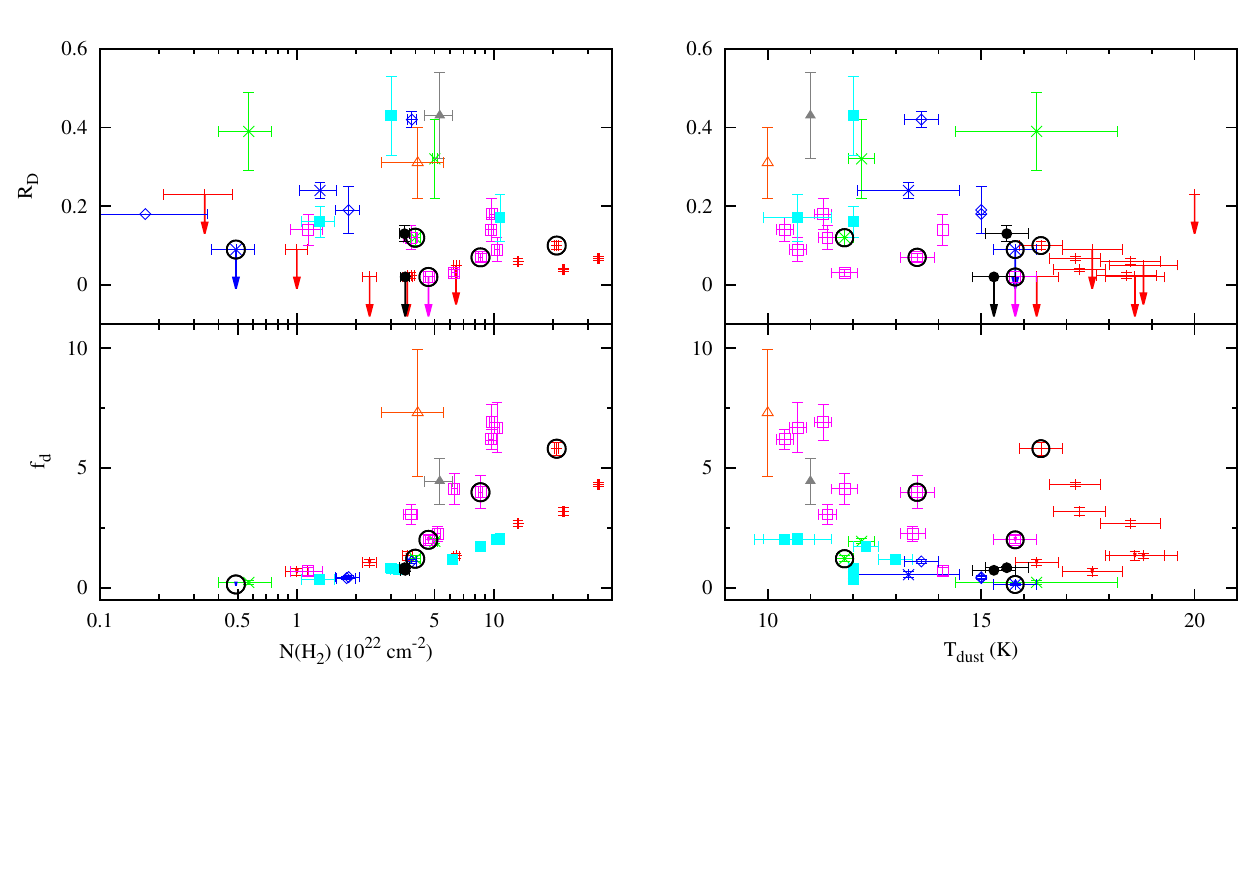}
\caption{Left upper panel: deuterium fraction depending on the molecular hydrogen column density, N(H$_2$). Right upper panel: deuterium fraction as a function of dust temperature taken from \citet{Pattle2015,Friesen2009} and \citet{Motte1998}. Left lower panel: CO-depletion factor versus the molecular hydrogen column density. Right panel: CO-depletion factor as a function of dust temperature.  The protostellar cores are depicted with black open circles, the color coding is given in Fig.~\ref{del-VLSR}.}\label{RD-flux}
\end{figure*}

The measured R$_{\rm D}$ and $f_d$ values are plotted as a function of molecular hydrogen column density, N(H$_2$), and dust temperature, T$_{\rm dust}$, in Fig.\,\ref{RD-flux}. No correlation is found for R$_{\rm D}$ vs N(H$_2$), while $f_d$ appears to increase with N(H$_2$) with different slopes depending on the region. In particular, $f_d$ is increasing faster with N(H$_2$) in the B2 region compared to the A and C regions. For the other regions, it is hard to see any trend, probably because of the more limited range of N(H$_2$) values detected. One possible cause of the different slopes in the $f_d$ -- N(H$_2$) correlations is the different amount of external heating due to the proximity of Oph-A to HD 147889 (see next section and Fig.~\ref{fig:RD-fd-dist}). This extra illumination maintains the dust grains in region A at a higher temperature compared to the other L1688 regions \citep[see also][]{Liseau2015}, so that larger column/volume densities are needed to reach dust temperatures low enough ($<$25\,K) to allow CO molecules to freeze-out onto the dust grains. The C region which rather follow the same trend as the A region is the next close to HD~147889 after the A region (see Fig.~\ref{fig:RD-fd-dist}) and probably also illuminated by the star, although it's temperature ($\le 15$~K) and column density ($\le 10^{22}$~cm$^{-2}$) are as low as in the other regions.

The difference in dust temperatures among the various regions in L1688 is also causing the scatter plot in the right panels of Fig.\,\ref{RD-flux}. Here we note that the different amount of external illumination impinging the different L1688 regions causes the well-known $f_d$ vs T$_{\rm dust}$ correlation \citep[e.g.][]{Kramer1999} to disappear. Oph-A is the only region in L1688 with dust temperatures larger than 15\,K and significant CO freeze-out.  Oph-B2 displays a sharp drop of $f_d$ with increasing T$_{\rm dust}$, as expected given the exponential dependence on T$_{\rm dust}$ of the CO evaporation rate \citep[see e.g.][]{Hasegawa1992}. 

The deuterium fraction is also expected (and has been measured) to drop with dust temperatures above about 15-20\,K \cite[.e.g.][]{Emprechtinger2009,Caselli2008,Kong2015}. However, this trend is not observed in L1688, as shown in the top left panel of Fig.\,\ref{RD-flux}. Once again, the non-uniform conditions among the various regions (in particular the amount of external illumination, gas volume density etc.) make it difficult to see a well defined pattern. Even within the same regions, we do not notice any trend, which may be caused by dust temperatures (mainly derived from Herschel data) not being representative of the cold regions within which the deuterium fractionation is taking place. More detailed and higher resolution data are needed to explore this possibility. 

\subsection{Distance to heating sources}
\label{section_heating}

\begin{figure*}\centering
\includegraphics[height=12cm,keepaspectratio]{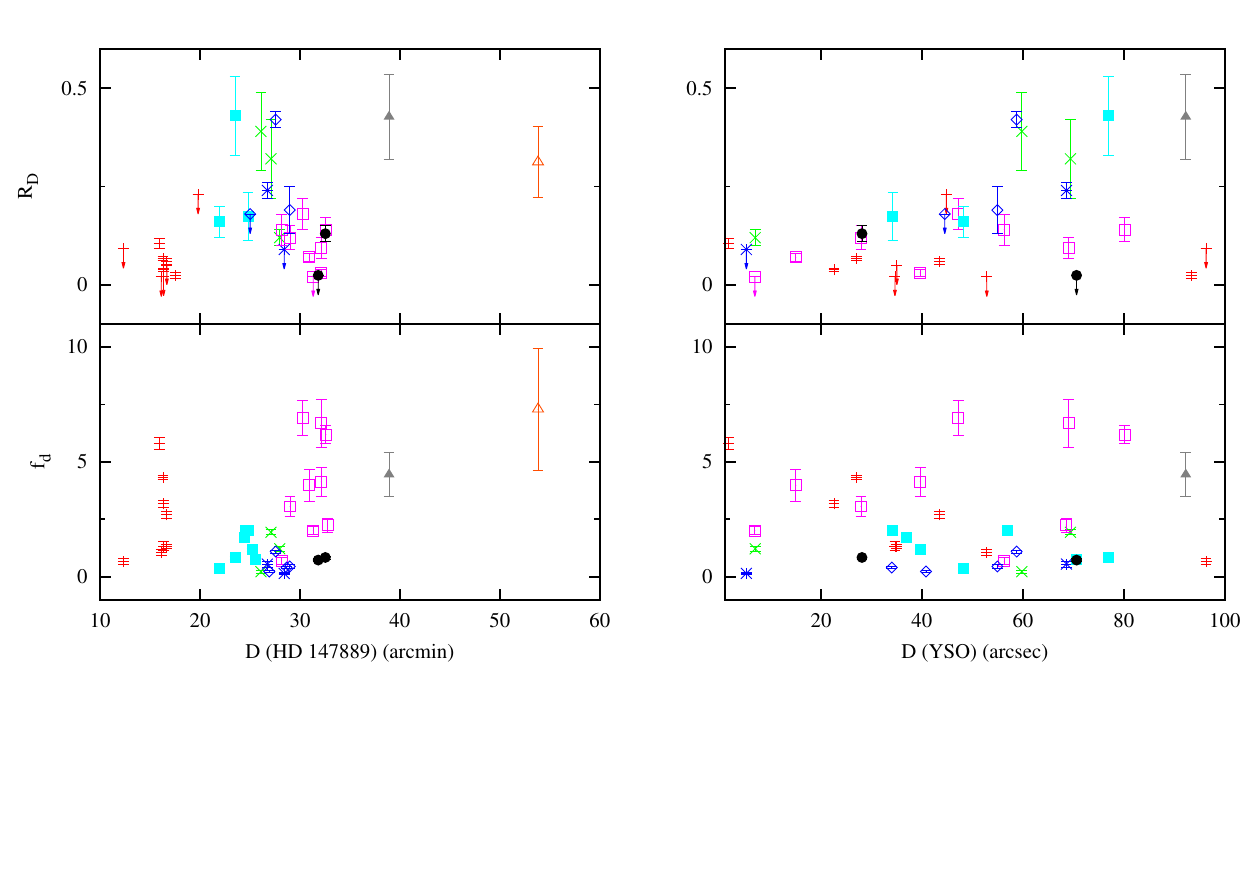}
\caption{Left upper panel: deuterium fraction depending on the angular distance to HD~147889. Right upper panel: deuterium fraction as a function of the angular distance to the closest YSO. The positions of YSOs were taken from \citet{Motte1998,Simpson2008} and \citet{Dunham2015}. Left lower pannel: CO-depletion factor depending on the angular distance to HD~147889. Right lower panel: CO-depletion factor depending on the angular distance to the closest YSO. The color coding is given in Fig.~\ref{del-VLSR}.}\label{fig:RD-fd-dist}
\end{figure*}

\begin{figure}\centering
\includegraphics[height=6cm,keepaspectratio]{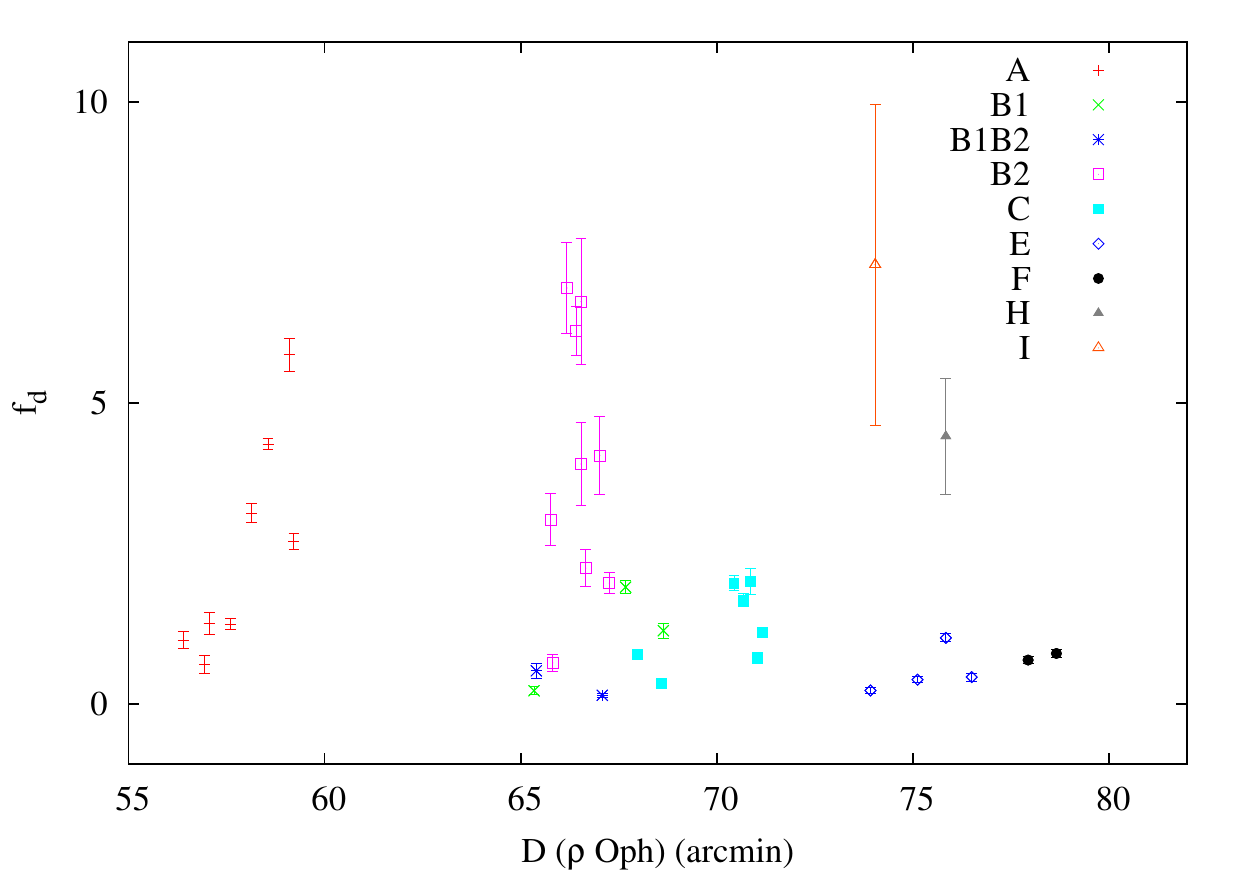}
\caption{CO-depletion factor as a function of angular distance to the $\rho$~Oph system.}\label{fd-rhoOph}
\end{figure}

It is well-known that star formation in L1688 is affected by the OB-association Sco~OB~2 \citep{Pattle2015}, which is located $\sim$11~pc behind L1688 \citep{Mamajek2008}. Sco~OB~2 is a moving group of more than 120 stars, mostly of B and A spectral types. It occupies an area of about 15$^{\circ}$ diameter on the sky \citep{Zeeuw1999}.  We looked for a correlation between R$_{\rm D}$, $f_d$, the distance to nearby stars ($\rho$~Oph, HD~147889, V~2246~Oph, Oph S1) and the closest YSOs (see Fig.~\ref{fig:RD-fd-dist}). $\rho$~Oph is a multiple system of B2IV--B2V spectral type stars; it is a member of the Sco~OB~2 association, located to the North of L1688, and is the most distant from L1688 among the four nearby stars. The cores in Oph-A, the nearest region to $\rho$~Oph, show a correlation between the CO-depletion factor and the projected angular distance D to the $\rho$~Oph system, with $f_d$ increasing with D (see Fig.~\ref{fd-rhoOph}). The cores in the others sub-regions do not show any correlation. 

The other three stars are pre-main sequence stars related to the Ophiuchus star-forming region. Oph S1 is a B4-K8 type binary star with a T-Tauri star being the fainter component \citep{Gagne2004} to the East of the A region; V~2246~Oph is a Herbig Ae/Be star on the West side of the A region; and HD~147889 is a pre-main sequence star on the West side of  L1688. In all regions of L1688, $f_d$ does not show any correlation with the distance to the brightest PMS stars of L1688. 

The A region contains more YSOs than other regions and is the closest to the external heating sources. Oph-A has the smallest values of R$_{\rm D}$ and the highest values of $f_d$ in L1688. Relatively low deuteration could be due to this proximity to nearby irradiating sources, while the large $f_d$ values may be due to the fact that this region is also the densest one in L1688, thus probably harbouring the densest cores. Here, the dust temperatures are probably low enough ($<$ 25\,K) to allow molecular freeze-out to proceed at a higher rate due to the larger gas-dust collision rates (because of the overall larger densities), but not large enough to promote deuterium fractionation, likely because of an increase of the ortho-to-para H$_2$ ratio in warmer environments \citep[see e.g.][]{Flower2006}. An alternative to the low values of R$_{\rm D}$ found in Oph-A could be that N$_2$D$^+$ cores are compact and small compared to our beam, while N$_2$H$^+$ is extended and abundant due to the large average densities \citep[see e.g.][]{Friesen2014,Liseau2015}, as also found in Infrared Dark Clouds where massive stars and star clusters form \citep{Henshaw2014}. Indeed, \citet{Friesen2014} detected compact (a few hundred AU in size) dust continuum condensation and ortho-H$_2$D$^+$ emission toward one of the cores embedded in Oph-A. However, we note that the line widths of ${\rm N_2H^+}$(1--0) and N$_2$D$^+$(1--0) lines are the same within the errors (see Fig.~\ref{fwhm}), so this alternative scenario may be harder to justify. Again, higher angular resolution observations of N$_2$D$^+$ are needed to disentangle between these two scenarios. 

No correlation between R$_{\rm D}$ and distance to any heating source (embedded or external) was found, although the largest R$_{\rm D}$ values are found at projected distances larger than 50\,arcsec from embedded YSOs and greater than 20\,arcsec from HD 147889 (these are the already discussed large R$_{\rm D}$ - low $f_d$ relatively isolated cores, which we discussed in section~\ref{sec:RD-fd}). Moreover, the region closest to $\rho$~Oph and V~2246~Oph, the sub-region Oph-A, has the lowest deuterium fractions.

\citet{Friesen2010b} studied the deuterium fraction in the B2 region and found a correlation between R$_{\rm D}$ and distance to the nearest protostar. The deuterium fraction we measure in this region (0--18\%, median $\sim$12\%) is systematically higher than that found by \citet{Friesen2010b} (0--10\%, mostly $<4$\%). According to our study, there is no correlation between R$_{\rm D}$ and the distance to the closest YSO (see Fig.~\ref{fig:RD-fd-dist} right upper panel). This difference in the results of the studies (R$_{\rm D}$ --- YSO correlation) can be due to the fewer cores studied by us in this particular region \citep[9 cores instead of full mapping of the B2 region as done by][]{Friesen2010b}.

%

\section{Conclusions}
This paper presents single point observations of the ground state transitions of ${\rm N_2H^+}$, ${\rm N_2D^+}$ and ${\rm C^{17}O}$ toward prestellar cores in L1688. We measure the deuterium fraction, ${\rm R_D}$, and CO-depletion factor, $f_d$, and study the correlation between these two parameters as well as with physical parameters varying across the cloud, such as dust temperature, molecular hydrogen column density, level of turbulence, projected distance to stellar sources which externally irradiate the cloud and embedded sources which can internally stir and shock the gas. The following conclusions have been reached:
   \begin{enumerate}
      \item The L1688 cores show a large spread of deuterium fractions, 2--43\% and moderate CO depletion factors (up to 7, although the reference value of the CO fractional abundance adopted here, $2\times10^{-4}$, may be underestimated by a factor of 2--3 or the dust opacity value used in the work, 0.01~cm$^2$~g$^{-1}$ may be overestimated by a factor of 1.5--2). 
      \item The largest R$_{\rm D}$ values are found toward (B1, B1B2, C, E, H, I) dense cores which present relatively quiescent (subsonic) motions as measured from the width of the high density tracer lines; thus, they are probably in an evolutionary stage just preceding the contraction toward protostellar birth, as found in other more evolved pre-stellar cores such as L1544 in Taurus \citep{Caselli2002-a}. These highly deuterated cores are also relatively isolated (they are typically found in between strong sub-millimetre dust continuum emission and far away from embedded protostars) and the CO freeze-out is low, in contrast with chemical model predictions. This dichotomy (large R$_{\rm D}$ and low $f_d$) can be understood if the deuterated gas is confined in a region smaller than the beam size, so that the CO-depleted region is too small to be revealed with observations of the widespread C$^{17}$O(1--0) emitting gas. Higher angular resolution observations are needed to confirm this statement. 
      \item Except for this sub-group of highly deuterated cores, widths are generally supersonic in C$^{17}$O(1--0) (78\%) and subsonic in N$_2$H$^+$(1--0) and N$_2$H$^+$(1--0) (75\% and 80\%). The B2 region stands out here with supersonic widths in all tracers in the majority of cores. This is probably a combination of internal systematic motions (e.g. contraction) and external stirring (turbulence, interaction with outflow driven by embedded protostars). 
      \item The correlation between ${\rm R_D}$ and $f_d$ already found in other studies of starless cores is maintained for a sub-group of cores (those in A, B2 and I, when plotted together). The highly deuterated cores show a significantly steeper rise of R$_{\rm D}$ with $f_d$, suggesting that the CO observations are not sensitive to the CO-depleted zone, as mentioned above.       
      \item The densest region in L1688, Oph-A, hosts dense cores with significant amount of CO freeze-out ($f_d$ close to 7) but no corresponding large R$_{\rm D}$ values.  This can be explained if the gas and dust temperatures are low enough ($<$25\,K) to allow CO freeze-out but high enough to significantly increase the ortho-to-para ratio compared to cooler regions. Alternatively, the N$_2$D$^+$ cores may be small and diluted within our beam, whereas the N$_2$H$^+$ is abundant all across the region as a consequence of the large average densities.  
      \item The nearby $\rho$ Oph star in the Sco OB2 association appears to affect the amount of CO freeze-out in Oph-A cores, as $f_d$ increases with projected distance to this star. No other regions appear to be chemically affected by their proximity to external stars or embedded young stellar objects, except for the B2 region, where $f_d$ decreases with increasing dust temperature. 
    \item Our observations hint at the present of compact starless cores (smaller than those found toward less dense and cooler molecular cloud complexes, such as Taurus) with large deuterium fractions (12--43\%) and small CO depletion factors (0.2--4.4). These cores have relatively low temperatures compared to their surroundings, so that they do not appear as clear structures in dust continuum emission maps at 850~$\mu m$. Their compact nature may be a consequence of the overall higher densities and temperatures across L1688 compared to other nearby star forming regions.  High angular resolution observations are needed to test these predictions. 
          
   \end{enumerate}

\begin{acknowledgements}
The authors thank J.~Pineda for thoughtful comments which helped to improve the manuscript. The authors acknowledge the financial support of the European Research Council (ERC; project PALs 320620); Andy Pon  acknowledge that partial salary support was provided by a CITA National Fellowship.
\end{acknowledgements}


\bibliography{Punanova_lit}{}
\bibliographystyle{aa}

\begin{appendix}
\include{./app}
\end{appendix}

\end{document}

%% file: app.tex
\section{Hyperfine splitting fit results}\label{app:tables}

Please note that in tables~\ref{N2H-plus-results}--\ref{CO21-results} the quantity labelled $T_{ant}\times\tau$, is $\tau\times(J_\nu(T_{ex})-J_\nu(T_{bg}))$ in the case of optically thick transition ($\tau>0.1$) or the main beam temperature ($T_{mb}$) in the case that the line is optically thin (with an adopted value of $\tau = 0.1$). For the details, see section~\ref{sec:N-RD}.

\begin{table*}
{\caption{Dense core coordinates and dataset numbers. Given coordinates are the observed positions, the centres of the cores determined by \citet{Motte1998}.}\label{lines}
\begin{tabular}{lccccccc}
\hline
\hline
\\
Core & $\alpha_{J2000}$ & $\delta_{J2000}$ & N$_2$H$^+$(1--0) &N$_2$D$^+$(1--0) & N$_2$D$^+$(2--1) & C$^{17}$O(1--0) & starless (s)/\\
   & ($^h$ $^m$ $^s$) & ($^{\circ}$ $^{\prime}$ $^{\prime\prime}$) & \multicolumn{4}{c}{Dataset}&protostellar (p)\\
\hline
A3-MM1	&	16:26:09.7	&	-24:23:06	& 188-97 &066-04&066-04& 188-97& s\\
A-MM4	&	16:26:24.1	&	-24:21:52	& 188-97 &066-04&066-04& 188-97& s\\
A-MM5	&	16:26:25.9	&	-24:22:27	& 188-97 &066-04&066-04& 188-97& s\\
VLA1623	&	16:26:26.5	&	-24:24:31	& 188-97 &066-04&066-04& 188-97& p\\
SM1N	&	16:26:27.3	&	-24:23:28	& 188-97&066-04&066-04& 188-97& s\\
SM1		&	16:26:27.5	&	-24:23:56	& 188-97&066-04&066-04& 188-97& s\\
A-MM6	&	16:26:27.9	&	-24:22:53	& 188-97&066-04&066-04& 188-97& s\\
SM2		&	16:26:29.5	&	-24:24:27	&066-04&066-04&066-04& 188-97& s\\
A-MM8	&	16:26:33.4	&	-24:25:01	&066-04&066-04&066-04& -- & s\\
A-S		&	16:26:43.1	&	-24:25:42	& 188-97&066-04&066-04& --  & s\\
\hline
B1-MM1	&	16:27:08.7	&	-24:27:50	& 051-00&066-04&066-04&066-04& s\\
B1-MM3	&	16:27:12.4	&	-24:29:58	& 188-97&066-04&066-04&066-04& s\\
B1-MM4	&	16:27:15.7	&	-24:30:42	& 188-97&066-04&066-04& 188-97& p\\
\hline
B1B2-MM1&	16:27:11.3	&	-24:27:39	& 051-00&066-04&066-04&066-04& s\\
B1B2-MM2e&	16:27:18.0	&	-24:28:48	& 051-00&066-04& -- &066-04& p\\
\hline
B2-MM1	&	16:27:17.0	&	-24:27:32	& 051-00&066-04&066-04&066-04& s\\
B2-MM2	&	16:27:20.3	&	-24:27:08	& 051-00&066-04&066-04&066-04& s\\
B2-MM6	&	16:27:25.3	&	-24:27:00	& 188-97&066-04&066-04&066-04& s\\
B2-MM8	&	16:27:28.0	&	-24:27:07	&066-04&066-04&066-04& 188-97& p\\
B2-MM10	&	16:27:29.6	&	-24:27:42	& 188-97&066-04&066-04&066-04& p\\
B2-MM14	&	16:27:32.8	&	-24:26:29	& 051-00&066-04&066-04& 051-00 & s\\
B2-MM15	&	16:27:32.8	&	-24:27:03	& 051-00&066-04&066-04&066-04& s\\
B2-MM16	&	16:27:34.5	&	-24:26:12	& 188-97&066-04&066-04& 051-00 & s\\
B2-MM17 &   16:27:35.2  &   -24:26:21   & 051-00& -- & -- & 051-00 & s\\
\hline
C-We	&	16:26:50.0	&	-24:32:49	& 051-00&066-04&066-04&066-04& s\\
C-Ne	&	16:26:57.2	&	-24:31:39	& 051-00&066-04&066-04& 051-00  & s\\
C-MM3   &   16:26:58.9  &   -24:34:22   & 051-00& -- & -- & 051-00  & s\\
C-MM4   &   16:26:59.4  &   -24:34:02   & 051-00& -- & -- & 051-00  & s\\
C-MM5	&	16:27:00.1	&	-24:34:27	& 188-97&066-04&066-04& 051-00  & s\\
C-MM6   &   16:27:01.6  &   -24:34:37   & 051-00& -- & -- & 051-00  & s\\
C-MM7   &   16:27:03.3  &   -24:34:22   & 051-00& -- & -- & 051-00  & s\\
\hline
E-MM1e	&	16:26:57.7	&	-24:36:56	& 051-00&066-04&066-04& -- & s\\
E-MM2d	&	16:27:04.9	&	-24:39:15	& 188-97&066-04&066-04& 051-00  & s\\
E-MM3   &   16:27:07.4  &   -24:36:01   & -- & -- & -- & 188-97& s\\
E-MM4	&	16:27:10.6	&	-24:39:30	&066-04&066-04&066-04& 051-00  & s\\
E-MM5	&	16:27:11.8	&	-24:37:57	& -- & -- & -- & 188-97& s\\
\hline
F-MM1	&	16:27:22.1	&	-24:40:02	&066-04&066-04&066-04& 188-97& s\\
F-MM2	&	16:27:24.3	&	-24:40:35	& 188-97&066-04&066-04& 188-97& s\\
\hline
H-MM1	&	16:27:58.3	&	-24:33:42	&066-04&066-04&066-04&066-04& s\\
I-MM1	&	16:28:57.7	&	-24:20:48	&066-04&066-04&066-04&066-04& s\\
\hline
\end{tabular}
}
\end{table*}

\begin{table*}
{\caption{Results of hfs-fitting of N$_2$H$^+$(1--0), excitation temperature T$_{ex}$ and total column density N$_{tot}$ calculations.}\label{N2H-plus-results}
\begin{tabular}{lcccccccc}
\hline
\hline
\\
Source & T$_{ant}\cdot\tau$       & V$_{\rm LSR}$         & $\Delta$v         & $\tau$ & RMS & T$_{ex}$ &N$_{tot}$\\
       & (K~km~s$^{-1}$) & (km~s$^{-1}$) & (km~s$^{-1}$) &        &T$_{mb}$ (K) & (K)       &($10^{13}$~cm$^{-2}$) \\
\hline

\hline
A3-MM1	 &   1.82 $\pm$   0.12 &   3.205 $\pm$   0.028 &    0.871 $\pm$   0.057 &     0.1 &   0.097 &    7.3$^a$  &    0.22 $\pm$    0.02 \\
A-MM4	 &  17.92 $\pm$   0.83 &   3.194 $\pm$   0.003 &    0.449 $\pm$   0.008 &     2.3 $\pm$     0.5 &   0.152 &   11.0 $\pm$    4.6 &    1.19 $\pm$    0.28 \\
A-MM5	 &  15.76 $\pm$   0.78 &   3.162 $\pm$   0.004 &    0.508 $\pm$   0.012 &     0.1 &   0.254 &    7.3  &    1.10 $\pm$    0.06 \\
VLA1623A &  20.79 $\pm$   0.52 &   3.635 $\pm$   0.002 &    0.561 $\pm$   0.005 &     4.5 $\pm$     0.3 &   0.105 &    7.7 $\pm$    0.9 &    1.60 $\pm$    0.19 \\
SM1N	 &  58.19 $\pm$   1.27 &   3.523 $\pm$   0.002 &    0.500 $\pm$   0.005 &     8.2 $\pm$     0.3 &   0.244 &   10.2 $\pm$    0.8 &    4.21 $\pm$    0.37 \\
SM1	 	 &  44.75 $\pm$   0.11 &   3.599 $\pm$   0.001 &    0.589 $\pm$   0.001 &     6.2 $\pm$     0.0 &   0.224 &   10.3 $\pm$    0.1 &    3.82 $\pm$    0.30 \\
A-MM6	 &   5.13 $\pm$   0.88 &   3.334 $\pm$   0.017 &    0.724 $\pm$   0.049 &     0.1 &   0.157 &    7.3  &    0.51 $\pm$    0.09 \\
SM2	 	 &  23.99 $\pm$   0.35 &   3.482 $\pm$   0.001 &    0.467 $\pm$   0.003 &     3.1 $\pm$     0.2 &   0.178 &   10.9 $\pm$    1.1 &    1.65 $\pm$    0.15 \\
A-MM8	 &  21.50 $\pm$   0.36 &   3.484 $\pm$   0.001 &    0.371 $\pm$   0.003 &     2.7 $\pm$     0.2 &   0.128 &   11.2 $\pm$    1.5 &    1.19 $\pm$    0.12 \\
A-S	 	 &   3.28 $\pm$   0.76 &   3.672 $\pm$   0.008 &    0.242 $\pm$   0.024 &     0.1 &   0.110 &    7.3  &    0.11 $\pm$    0.03 \\
\hline
B1-MM1	 &   6.10 $\pm$   1.30 &   4.017 $\pm$   0.009 &    0.291 $\pm$   0.024 &     0.1 &   0.186 &    7.3  &    0.24 $\pm$    0.06 \\
B1-MM3	 &  28.35 $\pm$   1.83 &   3.183 $\pm$   0.012 &    0.248 $\pm$   0.016 &    0.1 &   0.273 &    7.3  &    0.97 $\pm$    0.09 \\
	 	 &  20.72 $\pm$   2.09 &   3.775 $\pm$   0.006 &    0.333 $\pm$   0.014 &     4.3 $\pm$     1.3 &   0.273 &    7.9 $\pm$    4.3 &    0.95 $\pm$    0.30 \\
B1-MM4	 &  43.15 $\pm$   3.11 &   3.660 $\pm$   0.015 &    0.389 $\pm$   0.023 &    25.3 $\pm$     1.9 &   0.369 &    4.6 $\pm$    0.6 &    2.87 $\pm$    0.46 \\
	 	 &  13.62 $\pm$   2.41 &   3.985 $\pm$   0.006 &    0.245 $\pm$   0.014 &     0.1 &   0.369 &    7.3  &    0.46 $\pm$    0.09 \\
\hline
B1B2-MM1 &   9.56 $\pm$   0.57 &   4.031 $\pm$   0.005 &    0.385 $\pm$   0.013 &     0.1 &   0.192 &    7.3  &    0.51 $\pm$    0.03 \\
B1B2-MM2e&   3.84 $\pm$   0.98 &   3.913 $\pm$   0.026 &    0.570 $\pm$   0.071 &     0.1 &   0.188 &    7.3  &    0.30 $\pm$    0.09 \\
\hline
B2-MM1	 &  16.30 $\pm$   1.55 &   4.009 $\pm$   0.006 &    0.381 $\pm$   0.015 &     4.8 $\pm$     1.2 &   0.245 &    6.5 $\pm$    2.8 &    0.87 $\pm$    0.24 \\
B2-MM2	 &   9.33 $\pm$   0.82 &   3.841 $\pm$   0.023 &    0.628 $\pm$   0.059 &     3.9 $\pm$     0.9 &   0.121 &    5.4 $\pm$    1.9 &    0.89 $\pm$    0.25 \\
	 	 &   7.36 $\pm$   0.94 &   4.303 $\pm$   0.009 &    0.319 $\pm$   0.023 &    0.1 &   0.121 &    7.3  &    0.32 $\pm$    0.05 \\
B2-MM6	 &  19.64 $\pm$   1.61 &   3.724 $\pm$   0.010 &    0.696 $\pm$   0.026 &     4.0 $\pm$     0.8 &   0.303 &    8.0 $\pm$    3.0 &    1.88 $\pm$    0.43 \\
B2-MM8	 &   7.89 $\pm$   0.84 &   3.463 $\pm$   0.017 &    0.460 $\pm$   0.026 &    10.7 $\pm$     1.7 &   0.138 &    3.6 $\pm$    0.6 &    1.01 $\pm$    0.22 \\
	 	 &  23.63 $\pm$   0.43 &   4.133 $\pm$   0.003 &    0.507 $\pm$   0.006 &     3.8 $\pm$     0.2 &   0.138 &    9.3 $\pm$    1.0 &    1.69 $\pm$    0.17 \\
B2-MM10	 &  11.85 $\pm$   1.04 &   4.293 $\pm$   0.010 &    0.579 $\pm$   0.024 &     0.1 &   0.273 &    7.3  &    0.95 $\pm$    0.09 \\
B2-MM14	 &  17.13 $\pm$   1.78 &   4.122 $\pm$   0.014 &    0.771 $\pm$   0.035 &     3.7 $\pm$     1.0 &   0.236 &    7.7 $\pm$    3.8 &    1.82 $\pm$    0.53 \\ 
B2-MM15	 &  34.75 $\pm$   3.16 &   4.372 $\pm$   0.006 &    0.374 $\pm$   0.014 &     6.7 $\pm$     1.3 &   0.270 &    8.3 $\pm$    3.0 &    1.80 $\pm$    0.38 \\ 
B2-MM16	 &  17.27 $\pm$   1.14 &   4.003 $\pm$   0.008 &    0.640 $\pm$   0.017 &     3.0 $\pm$     0.7 &   0.232 &    8.8 $\pm$    3.6 &    1.54 $\pm$    0.37 \\ 
B2-MM17	 &  15.50 $\pm$   1.82 &   4.038 $\pm$   0.013 &    0.635 $\pm$   0.032 &     0.1 &   0.236 &    7.3  &    1.36 $\pm$    0.17 \\ 
\hline
C-We	 &  16.50 $\pm$   1.88 &   3.540 $\pm$   0.003 &    0.195 $\pm$   0.009 &     6.7 $\pm$     1.7 &   0.182 &    5.4 $\pm$    2.1 &    0.48 $\pm$    0.13 \\
C-Ne	 &  41.94 $\pm$   1.64 &   3.772 $\pm$   0.001 &    0.210 $\pm$   0.003 &    12.3 $\pm$     0.7 &   0.155 &    6.5 $\pm$    0.7 &    1.23 $\pm$    0.15 \\
C-MM3	 &  36.01 $\pm$   0.20 &   3.786 $\pm$   0.003 &    0.325 $\pm$   0.003 &    11.8 $\pm$     0.2 &   0.196 &    6.1 $\pm$    0.2 &    1.67 $\pm$    0.19 \\
C-MM4	 &  34.24 $\pm$   6.92 &   3.807 $\pm$   0.010 &    0.341 $\pm$   0.021 &    16.3 $\pm$     4.4 &   0.277 &    5.1 $\pm$    2.3 &    1.84 $\pm$    0.56 \\ 
C-MM5	 &  48.70 $\pm$   1.86 &   3.758 $\pm$   0.002 &    0.320 $\pm$   0.004 &    18.3 $\pm$     0.9 &   0.179 &    5.7 $\pm$    0.5 &    2.29 $\pm$    0.29 \\
C-MM6	 &  42.64 $\pm$   1.86 &   3.670 $\pm$   0.001 &    0.330 $\pm$   0.004 &    16.9 $\pm$     1.0 &   0.118 &    5.5 $\pm$    0.6 &    2.10 $\pm$    0.28 \\
C-MM7	 &  35.74 $\pm$   2.66 &   3.647 $\pm$   0.012 &    0.300 $\pm$   0.015 &    28.1 $\pm$     1.6 &   0.242 &    4.2 $\pm$    0.4 &    2.12 $\pm$    0.33 \\ 
\hline
E-MM1e	 &   2.51 $\pm$   0.20 &   4.396 $\pm$   0.019 &    0.361 $\pm$   0.028 &     0.1  &   0.181 &    7.3  &    0.12 $\pm$    0.01 \\
E-MM2d	 &  14.40 $\pm$   0.91 &   4.455 $\pm$   0.003 &    0.296 $\pm$   0.008 &     2.8 $\pm$     0.7 &   0.148 &    8.2 $\pm$    3.9 &    0.59 $\pm$    0.17$^b$ \\
E-MM3 \\
E-MM4	 &   7.29 $\pm$   0.42 &   4.197 $\pm$   0.003 &    0.283 $\pm$   0.007 &     2.2 $\pm$     0.7 &   0.120 &    6.3 $\pm$    3.0 &    0.29 $\pm$    0.09 \\
E-MM5 \\
\hline
F-MM1	 &  11.02 $\pm$   0.45 &   4.660 $\pm$   0.004 &    0.535 $\pm$   0.008 &     4.1 $\pm$     0.5 &   0.143 &    5.7 $\pm$    1.0 &    0.87 $\pm$    0.14 \\
F-MM2	 &  20.95 $\pm$   1.52 &   4.094 $\pm$   0.003 &    0.238 $\pm$   0.009 &     8.8 $\pm$     1.2 &   0.147 &    5.4 $\pm$    1.2 &    0.75 $\pm$    0.14 \\
		 &  9.20  $\pm$   0.74 &   4.550 $\pm$   0.005 &    0.340 $\pm$   0.013 &     0.1 &   0.147 &    7.3  &    0.43 $\pm$    0.04 \\
\hline
H-MM1	 & 23.02  $\pm$   0.47 &   4.223 $\pm$   0.001 &    0.271 $\pm$   0.002 &     3.7 $\pm$     0.3 &   0.209 &    9.3 $\pm$    1.2 &    0.88 $\pm$    0.09 \\
I-MM1	 &  28.57 $\pm$   0.74 &   3.296 $\pm$   0.001 &    0.271 $\pm$   0.003 &     8.9 $\pm$     0.4 &   0.142 &    6.3 $\pm$    0.5 &    1.09 $\pm$    0.13 \\
\hline
\end{tabular}\\
$^a$ In case when optical depth $\tau$ value was less than $3\sigma$ we did not calculate the excitation temperature with equation~\ref{Tex}, but rather adopt the average of the excitation temperatures of the other cores.\\
$^b$ N$_{tot}$ calculated assuming the line is optically thick, produce large uncertainty which propagates to the R$_{\rm D}$ uncertainty so R$_{\rm D}<2.5~\Delta{\rm R_D}$. Thus, to measure ${\rm R_D}$ for E-MM2d, N$_{tot}=(0.58\pm0.02)\times10^{13}$~$cm^{-2}$ calculated assuming the line is optically thin ($\tau=0.1$, T=7.3~K) used.
}
\end{table*}

\begin{table*}
{\caption{Results of hfs-fitting of N$_2$D$^+$(1--0)}\label{N2D-plus-results}
\begin{tabular}{lcccccccc}
\hline
\hline
Source & T$_{ant}\cdot\tau$       & V$_{\rm LSR}$         & $\Delta$v         & $\tau$ & RMS & T$_{ex}$ &N$_{tot}$\\
       & (K~km~s$^{-1}$) & (km~s$^{-1}$) & (km~s$^{-1}$) &        &T$_{mb}$ (K) & (K)       &($10^{12}$~cm$^{-2}$) \\
\hline
A3-MM1	 &&&&&      0.103 &     7.3 &       <0.20 \\  
A-MM4	 &&&&&      0.120 &    11.0 &       <0.26 \\
A-MM5	 &&&&&      0.120 &     7.3 &       <0.23 \\ 
VLA1623a &    1.56 $\pm$    0.09 &   3.644 $\pm$   0.017 &    0.546 $\pm$   0.032 &     0.1  &   0.114 &    7.7  &    1.57 $\pm$    0.13 \\ 
SM1N	 &    1.62 $\pm$    0.06 &   3.639 $\pm$   0.004 &    0.498 $\pm$   0.018 &     0.1  &   0.074 &   10.2  &    1.59 $\pm$    0.08 \\ 
SM1	 &    2.18 $\pm$    0.08 &   3.660 $\pm$   0.012 &    0.593 $\pm$   0.024 &     0.1  &   0.107 &   10.3  &    2.56 $\pm$    0.14 \\  
A-MM6	 &&&&&      0.136 &     7.3 &       <0.26 \\
SM2	 &    1.06 $\pm$    0.07 &   3.524 $\pm$   0.016 &    0.450 $\pm$   0.035 &     0.1  &   0.088 &   10.9  &    0.97 $\pm$    0.10 \\ 
A-MM8	 &    0.83 $\pm$    0.15 &   3.475 $\pm$   0.015 &    0.171 $\pm$   0.032 &     0.1  &   0.063 &   11.2  &    0.29 $\pm$    0.08 \\ 
A-S	 &&&&&      0.131 &     7.3 &       <0.25 \\ 
\hline
B1-MM1	 &    1.29 $\pm$    0.08 &   3.990 $\pm$   0.012 &    0.392 $\pm$   0.027 &     0.1  &   0.087 &    7.3  &    0.93 $\pm$    0.09 \\ 
B1-MM3	 &    4.71 $\pm$    0.43 &   3.855 $\pm$   0.006 &    0.371 $\pm$   0.014 &     6.5 $\pm$     1.3 &   0.090 &    3.5 $\pm$    0.5 &    6.22 $\pm$    1.62 \\ 
B1-MM4	 &    3.19 $\pm$    0.17 &   3.567 $\pm$   0.008 &    0.259 $\pm$   0.019 &     0.1  &   0.115 &    4.6  &    1.81 $\pm$    0.16 \\ 
	 &    2.94 $\pm$    0.15 &   3.995 $\pm$   0.010 &    0.363 $\pm$   0.029 &     0.1  &   0.115 &    4.6  &    2.34 $\pm$    0.22 \\ 
\hline
B1B2-MM1 &    1.76 $\pm$    0.08 &   4.021 $\pm$   0.008 &    0.383 $\pm$   0.018 &     0.1  &   0.086 &    7.3  &    1.24 $\pm$    0.08 \\
B1B2-MM2 &&&&&      0.135 &     7.3 &       <0.26 \\ 
\hline 
B2-MM1	 &    1.77 $\pm$    0.09 &   4.082 $\pm$   0.008 &    0.374 $\pm$   0.021 &     0.1  &   0.092 &    6.5  &    1.22 $\pm$    0.09 \\ 
B2-MM2	 &    1.00 $\pm$    0.07 &   4.194 $\pm$   0.029 &    0.752 $\pm$   0.053 &     0.1  &   0.100 &    5.4  &    1.46 $\pm$    0.15 \\ 
B2-MM6	 &    2.38 $\pm$    0.07 &   3.840 $\pm$   0.013 &    0.759 $\pm$   0.024 &     0.1  &   0.107 &    8.0  &    3.34 $\pm$    0.15 \\
B2-MM8	 &    0.78 $\pm$    0.06 &   3.901 $\pm$   0.048 &    1.241 $\pm$   0.114 &     0.1  &   0.104 &    9.3  &    1.85 $\pm$    0.23 \\ 
B2-MM10	 &&&&&      0.119 &     7.3 &       <0.23 \\ 
B2-MM14	 &    1.46 $\pm$    0.10 &   4.129 $\pm$   0.035 &    0.622 $\pm$   0.049 &     0.1  &   0.129 &    7.0  &    1.67 $\pm$    0.17 \\  
B2-MM15	 &    0.95 $\pm$    0.10 &   4.365 $\pm$   0.018 &    0.350 $\pm$   0.045 &     0.1  &   0.101 &    7.3  &    0.61 $\pm$    0.10 \\ 
B2-MM16	 &    1.85 $\pm$    0.08 &   3.877 $\pm$   0.016 &    0.606 $\pm$   0.019 &     0.1  &   0.127 &    8.8  &    2.11 $\pm$    0.11 \\ 
B2-MM17 \\
\hline
C-We	 &    1.63 $\pm$    0.10 &   3.575 $\pm$   0.007 &    0.236 $\pm$   0.015 &     0.1  &   0.088 &    5.4  &    0.75 $\pm$    0.07 \\ 
C-N	 &   10.07 $\pm$    0.65 &   3.795 $\pm$   0.002 &    0.228 $\pm$   0.006 &     6.4 $\pm$     0.9 &   0.127 &    4.4 $\pm$    0.7 &    5.25 $\pm$    1.08 \\ 
C-MM3 \\
C-MM4 \\
C-MM5	 &    7.42 $\pm$    0.49 &   3.881 $\pm$   0.003 &    0.282 $\pm$   0.008 &     2.5 $\pm$     0.8 &   0.127 &    5.9 $\pm$    2.4 &    3.93 $\pm$    1.29 \\ 
C-MM6 \\
C-MM7 \\
\hline
E-MM1	 &&&&&      0.113 &     7.3 &       <0.22 \\ 
E-MM2d	 &    5.91 $\pm$    0.53 &   4.485 $\pm$   0.004 &    0.295 $\pm$   0.011 &     4.9 $\pm$     1.2 &   0.115 &    4.0 $\pm$    1.0 &    4.51 $\pm$    1.30$^a$ \\
E-MM3 \\
E-MM4	 &    0.82 $\pm$    0.07 &   4.288 $\pm$   0.016 &    0.357 $\pm$   0.033 &     0.1  &   0.081 &    6.3  &    0.55 $\pm$    0.07 \\ 
E-MM5 \\
\hline
F-MM1	 &&&&&      0.087 &     5.7 &       <0.18 \\  
F-MM2	 &    1.03 $\pm$    0.09 &   4.276 $\pm$   0.033 &    0.767 $\pm$   0.069 &     0.1  &   0.121 &    5.4  &    1.54 $\pm$    0.19 \\ 
\hline
H-MM1	 &    6.95 $\pm$    0.30 &   4.243 $\pm$   0.002 &    0.280 $\pm$   0.005 &     2.7 $\pm$     0.5 &   0.081 &    5.5 $\pm$    1.4 &    3.76 $\pm$    0.86 \\ 
I-MM1 	 &    5.88 $\pm$    0.31 &   3.325 $\pm$   0.002 &    0.283 $\pm$   0.006 &     2.8 $\pm$     0.6 &   0.079 &    5.0 $\pm$    1.4 &    3.40 $\pm$    0.89 \\ 
\hline 
\end{tabular}\\
$^a$ N$_{tot}$ calculated assuming the line is optically thick, produce large uncertainty which propagates to the R$_{\rm D}$ uncertainty so R$_{\rm D}<2.5~\Delta{\rm R_D}$. Thus, to measure ${\rm R_D}$ for E-MM2d, N$_{tot}=(2.41\pm0.07)\times10^{12}$~$cm^{-2}$ calculated assuming the line is optically thin ($\tau=0.1$, T=7.3~K) used.
} 
\end{table*}

\begin{table*}
{\caption{Results of hfs-fitting of N$_2$D$^+$(2--1)}\label{N2D-plus-results-21}
\begin{tabular}{lcccccccc}
\hline
\hline
Source & T$_{ant}\cdot\tau$       & V$_{\rm LSR}$         & $\Delta$v         & $\tau$ & RMS & T$_{ex}$ &N$_{tot}$\\
       & (K~km~s$^{-1}$) & (km~s$^{-1}$) & (km~s$^{-1}$) &        &T$_{mb}$ (K) & (K)       &($10^{12}$~cm$^{-2}$) \\
\hline
A3-MM1	 &&&&&  0.186 &     7.3 &	     <0.22 \\ 
A-MM4	 &    1.10 $\pm$    0.15 &   3.169 $\pm$   0.019 &    0.303 $\pm$   0.047 &	0.1  &   0.170 &   11.0  &    0.31 $\pm$    0.06 \\ 
A-MM5	 &&&&&  0.275 &     7.3 &	     <0.32 \\ 
VLA1623A &    4.13 $\pm$    0.08 &   3.653 $\pm$   0.006 &    0.549 $\pm$   0.009 &	0.1  &   0.268 &    7.7  &    2.43 $\pm$    0.06 \\ 
SM1N	 &    5.42 $\pm$    0.25 &   3.627 $\pm$   0.004 &    0.468 $\pm$   0.011 &	1.2 $\pm$     0.2 &   0.139 &	 8.0 $\pm$    3.3 &    2.64 $\pm$    0.45 \\ 
SM1	 &    3.38 $\pm$    0.37 &   3.754 $\pm$   0.011 &    0.493 $\pm$   0.028 &	1.6 $\pm$     0.5 &   0.202 &	 5.6 $\pm$    3.9 &    2.54 $\pm$    0.83 \\ 
A-MM6	 &&&&&  0.183 &     7.3 &	     <0.22 \\ 
SM2	 &    2.74 $\pm$    0.15 &   3.435 $\pm$   0.008 &    0.311 $\pm$   0.019 &	0.1  &   0.193 &   10.9  &    0.79 $\pm$    0.06 \\ 
A-MM8	 &    1.42 $\pm$    0.23 &   3.411 $\pm$   0.035 &    0.383 $\pm$   0.068 &	0.1  &   0.223 &   11.2  &    0.50 $\pm$    0.12 \\ 
A-S	 &&&&&  0.178 &     7.3 &	     <0.21 \\ 
\hline
B1-MM1   &    1.84 $\pm$    0.32 &   3.952 $\pm$   0.023 &    0.298 $\pm$   0.060 &	0.1  &   0.215 &    7.3  &    0.61 $\pm$    0.16 \\ 
B1-MM3   &    3.53 $\pm$    0.29 &   3.747 $\pm$   0.011 &    0.321 $\pm$   0.032 &	0.1  &   0.365 &    7.9  &    1.19 $\pm$    0.15 \\ 
B1-MM4   &    1.85 $\pm$    0.17 &   3.752 $\pm$   0.036 &    0.702 $\pm$   0.066 &	0.1  &   0.309 &    4.6  &    2.84 $\pm$    0.37 \\ 
\hline
B1B2-MM1 &    2.11 $\pm$    0.27 &   3.978 $\pm$   0.016 &    0.274 $\pm$   0.041 &	0.1  &   0.310 &    7.3  &    0.65 $\pm$    0.13 \\ 
B1B2-MM2e \\
\hline
B2-MM1   &    3.23 $\pm$    0.21 &   3.967 $\pm$   0.010 &    0.286 $\pm$   0.019 &	0.1  &   0.294 &    6.5  &    1.15 $\pm$    0.11 \\ 
B2-MM2	 &    1.73 $\pm$    0.14 &   4.072 $\pm$   0.030 &    0.743 $\pm$   0.070 &     0.1  &   0.192 &    5.4  &    2.06 $\pm$    0.26 \\ 
B2-MM6   &    3.90 $\pm$    0.14 &   3.785 $\pm$   0.013 &    0.735 $\pm$   0.031 &	0.1  &   0.335 &    8.0  &    2.99 $\pm$    0.17 \\ 
B2-MM8   &    1.83 $\pm$    0.14 &   4.160 $\pm$   0.017 &    0.435 $\pm$   0.036 &	0.1  &   0.253 &    9.3  &    0.77 $\pm$    0.09 \\ 
B2-MM10	 &&&&&  0.326 &     7.3 &	     <0.38 \\ 
B2-MM14  &    3.97 $\pm$    0.11 &   4.160 $\pm$   0.006 &    0.426 $\pm$   0.014 &	0.1  &   0.197 &    7.0  &    1.96 $\pm$    0.08 \\ 
B2-MM15	 &    1.00 $\pm$    0.14 &   4.305 $\pm$   0.029 &    0.446 $\pm$   0.076 &     0.1  &   0.107 &    7.3  &    0.50 $\pm$    0.11 \\ 
B2-MM16  &    4.59 $\pm$    0.13 &   3.921 $\pm$   0.008 &    0.611 $\pm$   0.021 &	0.1  &   0.266 &    8.8  &    2.77 $\pm$    0.13 \\ 
B2-MM17 \\
\hline
C-W	 &    1.40 $\pm$    0.27 &   3.507 $\pm$   0.018 &    0.239 $\pm$   0.057 &	0.1  &   0.160 &    5.4  &    0.54 $\pm$    0.17 \\ 
C-N	 &   12.80 $\pm$    0.93 &   3.743 $\pm$   0.004 &    0.175 $\pm$   0.009 &	6.4 $\pm$     0.7 &   0.235 &	 5.4 $\pm$    1.4 &    3.59 $\pm$    0.46 \\ 
C-MM3 \\
C-MM4 \\
C-MM5	 &    7.83 $\pm$    0.98 &   3.829 $\pm$   0.006 &    0.222 $\pm$   0.018 &	4.2 $\pm$     0.9 &   0.270 &	 5.2 $\pm$    2.5 &    2.96 $\pm$    0.67 \\ 
C-MM6 \\
C-MM7 \\
\hline
E-MM1	 &&&&&  0.222 &     7.3 &	     <0.26 \\ 
E-MM2D   &   12.07 $\pm$    0.80 &   4.413 $\pm$   0.004 &    0.232 $\pm$   0.010 &	3.2 $\pm$     0.4 &   0.247 &	 7.3 $\pm$    2.5 &    3.11 $\pm$    0.43 \\ 
E-MM3 \\
E-MM4	 &    1.06 $\pm$    0.20 &   4.187 $\pm$   0.026 &    0.295 $\pm$   0.061 &	0.1  &   0.168 &	 6.3  &    0.41 $\pm$	 0.11 \\ 
E-MM5 \\
\hline
F-MM1	 &&&&&  0.230 &     5.7 &	     <0.36 \\ 
F-MM2	 &    1.21 $\pm$    0.10 &   4.244 $\pm$   0.032 &    0.948 $\pm$   0.100 &     0.1  &   0.197 &    5.4  &    1.83 $\pm$    0.25 \\ 
\hline
H-MM1	 &   19.87 $\pm$    0.74 &   4.208 $\pm$   0.002 &    0.217 $\pm$   0.005 &	4.4 $\pm$     0.3 &   0.209 &	 8.1 $\pm$    1.4 &    4.44 $\pm$    0.31 \\ 
I-MM1	 &    5.49 $\pm$    0.55 &   3.242 $\pm$   0.005 &    0.259 $\pm$   0.015 &	2.0 $\pm$     0.5 &   0.194 &	 6.3 $\pm$    3.9 &    1.84 $\pm$    0.51 \\ 
\hline
\end{tabular}\\
}
\end{table*}

\begin{table*}
{\caption{Results of hfs-fitting of C$^{17}$O(1--0) and column density calculations}\label{CO-results}
\begin{tabular}{lcccccccc}
\hline
\hline
Source & T$_{ant}\cdot\tau$       & V$_{\rm LSR}$         & $\Delta$v         & $\tau$ & RMS & T$_{ex}$ &N$_{tot}$\\
       & (K~km~s$^{-1}$) & (km~s$^{-1}$) & (km~s$^{-1}$) &        &T$_{mb}$ (K) & (K)       &($10^{15}$~cm$^{-2}$) \\
\hline
A3-MM1	 &    0.61 $\pm$    0.08 &   3.050 $\pm$   0.123 &    1.678 $\pm$   0.223 &	0.1  &   0.154 &   17.6  &    1.32 $\pm$    0.24 \\ 
A-MM4	 &    1.88 $\pm$    0.12 &   3.152 $\pm$   0.027 &    0.829 $\pm$   0.066 &	0.1  &   0.198 &   16.3  &    1.92 $\pm$    0.20 \\ 
A-MM5	 &    2.65 $\pm$    0.21 &   3.186 $\pm$   0.019 &    0.673 $\pm$   0.069 &	0.1  &   0.210 &   18.6  &    2.37 $\pm$    0.30 \\ 
VLA1623a &    3.35 $\pm$    0.09 &   3.714 $\pm$   0.007 &    0.755 $\pm$   0.025 &	0.1  &   0.114 &   16.4  &    3.13 $\pm$    0.13 \\ 
SM1N	 &    5.98 $\pm$    0.16 &   3.524 $\pm$   0.006 &    0.540 $\pm$   0.015 &	0.1  &   0.152 &   17.3  &    4.11 $\pm$    0.16 \\ 
		 &    0.90 $\pm$    0.10 &   3.096 $\pm$   0.097 &    1.812 $\pm$   0.111 &	0.1  &   0.152 &   17.3  &    2.09 $\pm$    0.27 \\ 
SM1		 &    7.77 $\pm$    0.09 &   3.659 $\pm$   0.004 &    0.696 $\pm$   0.010 &	0.1  &   0.134 &   17.2  &    6.86 $\pm$    0.13 \\ 
A-MM6	 &    3.30 $\pm$    0.11 &   3.094 $\pm$   0.013 &    0.961 $\pm$   0.040 &	0.1  &   0.209 &   18.8  &    4.24 $\pm$    0.23 \\ 
SM2		 &    4.06 $\pm$    0.12 &   3.492 $\pm$   0.010 &    0.796 $\pm$   0.029 &	0.1  &   0.159 &   18.5  &    4.28 $\pm$    0.20 \\ 
\hline
B1-MM1	 &    1.60 $\pm$    0.06 &   3.625 $\pm$   0.020 &    1.158 $\pm$   0.045 &	0.1  &   0.192 &   16.3  &    2.29 $\pm$    0.12 \\ 
B1-MM3	 &    2.12 $\pm$    0.07 &   3.480 $\pm$   0.015 &    0.984 $\pm$   0.038 &	0.1  &   0.202 &   16.4  &    2.58 $\pm$    0.13 \\ 
B1-MM4	 &    2.61 $\pm$    0.13 &   3.542 $\pm$   0.024 &    1.030 $\pm$   0.067 &	0.1  &   0.223 &   13.0  &    2.97 $\pm$    0.25 \\ 
\hline
B1B2-MM1 &    1.33 $\pm$    0.05 &   3.696 $\pm$   0.026 &    1.403 $\pm$   0.060 &	0.1  &   0.204 &   13.3  &    2.09 $\pm$    0.12 \\ 
B1B2-MM2 &    1.65 $\pm$    0.05 &   4.127 $\pm$   0.020 &    1.537 $\pm$   0.049 &	0.1  &   0.206 &   15.8  &    3.08 $\pm$    0.13 \\ 
\hline
B2-MM1	 &    1.70 $\pm$    0.11 &   3.772 $\pm$   0.021 &    0.752 $\pm$   0.060 &	0.1  &   0.268 &   14.1  &    1.46 $\pm$    0.15 \\ 
B2-MM2	 &    0.97 $\pm$    0.07 &   3.915 $\pm$   0.033 &    1.054 $\pm$   0.085 &	0.1  &   0.231 &   14.2  &    1.18 $\pm$    0.13 \\ 
B2-MM6	 &    0.81 $\pm$    0.06 &   3.362 $\pm$   0.054 &    1.444 $\pm$   0.112 &	0.1  &   0.246 &   13.2  &    1.30 $\pm$    0.14 \\ 
B2-MM8	 &    0.79 $\pm$    0.09 &   3.402 $\pm$   0.137 &    2.124 $\pm$   0.259 &	0.1  &   0.171 &   14.5  &    1.94 $\pm$    0.33 \\ 
B2-MM10  &    0.86 $\pm$    0.05 &   3.427 $\pm$   0.059 &    1.944 $\pm$   0.104 &	0.1  &   0.231 &   14.0  &    1.90 $\pm$    0.15 \\ 
B2-MM14  &    0.84 $\pm$    0.09 &   3.602 $\pm$   0.093 &    1.563 $\pm$   0.173 &	0.1  &   0.172 &   14.6  &    1.52 $\pm$    0.24 \\ 
B2-MM15  &    1.10 $\pm$    0.10 &   3.370 $\pm$   0.040 &    1.135 $\pm$   0.143 &	0.1  &   0.235 &   13.9  &    1.42 $\pm$    0.22 \\ 
B2-MM16  &    1.24 $\pm$    0.05 &   3.632 $\pm$   0.021 &    1.067 $\pm$   0.049 &	0.1  &   0.102 &   14.3  &    1.53 $\pm$    0.09 \\ 
B2-MM17  &    1.32 $\pm$    0.12 &   3.636 $\pm$   0.062 &    1.353 $\pm$   0.125 &	0.1  &   0.180 &   14.3  &    2.06 $\pm$    0.26 \\ 
\hline
C-We	 &    3.39 $\pm$    0.06 &   3.715 $\pm$   0.008 &    0.688 $\pm$   0.016 &	0.1  &   0.217 &   12.0  &    2.50 $\pm$    0.07 \\ 
		 &    2.07 $\pm$    0.09 &   4.395 $\pm$   0.010 &    0.398 $\pm$   0.016 &	0.1  &   0.217 &   12.0  &    0.88 $\pm$    0.05 \\ 
C-Ne	 &    3.82 $\pm$    0.11 &   3.744 $\pm$   0.009 &    0.782 $\pm$   0.029 &	0.1  &   0.153 &   12.0  &    3.21 $\pm$    0.15 \\ 
C-MM3	 &    6.37 $\pm$    0.21 &   3.716 $\pm$   0.009 &    0.626 $\pm$   0.026 &	0.1  &   0.228 &   12.3  &    4.32 $\pm$    0.23 \\ 
C-MM4	 &    6.10 $\pm$    0.21 &   3.721 $\pm$   0.009 &    0.598 $\pm$   0.026 &	0.1  &   0.212 &   10.4  &    3.75 $\pm$    0.21 \\ 
		 &    1.72 $\pm$    0.21 &   4.911 $\pm$   0.026 &    0.413 $\pm$   0.056 &	0.1  &   0.212 &   10.4  &    0.73 $\pm$    0.13 \\ 
C-MM5	 &    5.15 $\pm$    0.26 &   3.692 $\pm$   0.008 &    0.423 $\pm$   0.026 &	0.1  &   0.227 &   10.7  &    2.26 $\pm$    0.18 \\ 
		 &    1.05 $\pm$    0.18 &   3.962 $\pm$   0.114 &    2.140 $\pm$   0.225 &	0.1  &   0.227 &   10.7  &    2.33 $\pm$    0.46 \\ 
C-MM6	 &    5.07 $\pm$    0.15 &   3.676 $\pm$   0.008 &    0.709 $\pm$   0.028 &	0.1  &   0.173 &   13.0  &    3.98 $\pm$    0.20 \\ 
		 &    1.93 $\pm$    0.18 &   4.855 $\pm$   0.012 &    0.266 $\pm$   0.031 &	0.1  &   0.173 &   13.0  &    0.57 $\pm$    0.09 \\ 
C-MM7	 &    5.94 $\pm$    0.28 &   3.691 $\pm$   0.012 &    0.591 $\pm$   0.033 &	0.1  &   0.307 &   12.0  &    3.77 $\pm$    0.27 \\ 
\hline
E-MM2d	 &    1.27 $\pm$    0.03 &   3.447 $\pm$   0.016 &    1.128 $\pm$   0.029 &	0.1  &   0.080 &   13.6  &    1.62 $\pm$    0.05 \\ 
		 &    1.90 $\pm$    0.07 &   3.930 $\pm$   0.005 &    0.243 $\pm$   0.009 &	0.1  &   0.080 &   13.6  &    0.52 $\pm$    0.03 \\ 
		 &    1.29 $\pm$    0.04 &   4.502 $\pm$   0.014 &    0.623 $\pm$   0.022 &	0.1  &   0.080 &   13.6  &    0.91 $\pm$    0.04 \\ 
E-MM3	 &    1.02 $\pm$    0.15 &   3.700 $\pm$   0.099 &    2.261 $\pm$   0.184 &	0.1  &   0.175 &   15.0  &    2.71 $\pm$    0.46 \\ 
		 &    2.09 $\pm$    0.20 &   3.477 $\pm$   0.022 &    0.600 $\pm$   0.066 &	0.1  &   0.175 &   15.0  &    1.48 $\pm$    0.21 \\ 
E-MM4	 &    2.04 $\pm$    0.04 &   3.427 $\pm$   0.022 &    1.369 $\pm$   0.042 &	0.1  &   0.108 &   15.0  &    3.30 $\pm$    0.12 \\ 
		 &    1.16 $\pm$    0.14 &   4.197 $\pm$   0.011 &    0.216 $\pm$   0.035 &	0.1  &   0.108 &   15.0  &    0.30 $\pm$    0.06 \\ 
E-MM5	 &    2.42 $\pm$    0.11 &   3.314 $\pm$   0.023 &    0.952 $\pm$   0.065 &	0.1  &   0.176 &   15.0  &    2.72 $\pm$    0.22 \\ 
		 &    1.63 $\pm$    0.16 &   4.549 $\pm$   0.031 &    0.603 $\pm$   0.083 &	0.1  &   0.176 &   15.0  &    1.16 $\pm$    0.20 \\ 
\hline
F-MM1	 &    2.54 $\pm$    0.09 &   3.690 $\pm$   0.024 &    1.411 $\pm$   0.057 &	0.1  &   0.217 &   15.3  &    4.27 $\pm$    0.23 \\ 
F-MM2	 &    2.49 $\pm$    0.10 &   3.725 $\pm$   0.023 &    1.230 $\pm$   0.055 &	0.1  &   0.206 &   15.6  &    3.68 $\pm$    0.22 \\ 
\hline
H-MM1	 &    1.38 $\pm$    0.12 &   3.957 $\pm$   0.029 &    0.726 $\pm$   0.083 &	0.1  &   0.213 &   11.0  &    1.04 $\pm$    0.15 \\ 
I-MM1	 &    1.74 $\pm$    0.13 &   3.231 $\pm$   0.010 &    0.277 $\pm$   0.025 &	0.1  &   0.206 &   10.0  &    0.49 $\pm$    0.06 \\ 
\hline
\end{tabular}
}
\end{table*}

\begin{table*}
\caption{Results of hfs-fitting of C$^{17}$O(2--1) and column density calculations}\label{CO21-results}
\begin{tabular}{lccccccc}
\hline
\hline
\\
Source & T$_{ant}\cdot\tau$ & V$_{\rm LSR}$         & $\Delta$v  			& $\tau$      	& RMS	& T$_{ex}$&	N$_{tot}$\\
       & (K~km~s$^{-1}$) 	& (km~s$^{-1}$) 	 & (km~s$^{-1}$)			&				& (K)	& (K)	& (10$^{15}$~cm$^{-2}$)\\
\hline
A3-MM1	 &    3.94 $\pm$    0.55 &   2.513 $\pm$   0.051 &    0.702 $\pm$   0.110 &	0.1  &   0.846 &   17.6  &    1.57 $\pm$	0.33 \\ 
A-MM4	 &    6.85 $\pm$    0.62 &   3.231 $\pm$   0.034 &    0.682 $\pm$   0.067 &	0.1  &   0.966 &   16.3  &    2.66 $\pm$	0.36 \\ 
	 &    1.64 $\pm$    0.79 &   5.071 $\pm$   0.115 &    0.523 $\pm$   0.289 &	0.1  &   0.966 &   16.3  &    0.49 $\pm$	0.36 \\ 
A-MM5	 &    9.12 $\pm$    0.70 &   3.208 $\pm$   0.025 &    0.634 $\pm$   0.052 &	0.1  &   1.068 &   18.6  &    3.29 $\pm$	0.37 \\ 
VLA1623a  &    9.53 $\pm$    0.40 &   3.733 $\pm$   0.017 &    0.820 $\pm$   0.039 &	0.1  &   0.401 &   16.4  &    4.46 $\pm$	0.28 \\ 
SM1N	 &   15.79 $\pm$    0.47 &   3.496 $\pm$   0.006 &    0.469 $\pm$   0.018 &	0.1  &   0.308 &   17.3  &    4.21 $\pm$	0.21 \\ 
	 &    4.53 $\pm$    0.28 &   3.011 $\pm$   0.050 &    1.834 $\pm$   0.078 &	0.1  &   0.308 &   17.3  &    4.72 $\pm$	0.35 \\ 
SM1	 &   35.26 $\pm$    1.71 &   3.667 $\pm$   0.006 &    0.638 $\pm$   0.015 &	1.8 $\pm$ 0.2  &   0.572 &   17.2  &    7.95 $\pm$	0.92 \\ 
A-MM6	 &   12.29 $\pm$    0.54 &   3.038 $\pm$   0.021 &    0.942 $\pm$   0.046 &	0.1  &   0.966 &   18.8  &    6.60 $\pm$	0.43 \\ 
SM2	 &   10.29 $\pm$    0.67 &   3.322 $\pm$   0.030 &    0.937 $\pm$   0.045 &	0.1  &   0.699 &   18.5  &    5.49 $\pm$	0.44 \\ 
	 &    8.54 $\pm$    0.99 &   3.615 $\pm$   0.015 &    0.274 $\pm$   0.042 &	0.1  &   0.699 &   18.5  &    1.33 $\pm$	0.26 \\ 
A-MM8	 &    7.62 $\pm$    1.18 &   3.446 $\pm$   0.034 &    0.414 $\pm$   0.065 &	0.1  &   0.950 &   18.4  &    1.80 $\pm$	0.40 \\ 
\hline
B1-MM1	 &    4.04 $\pm$    0.43 &   3.611 $\pm$   0.061 &    1.235 $\pm$   0.155 &	0.1  &   0.516 &   16.3  &    2.84 $\pm$	0.47 \\ 
B1-MM4	 &    5.65 $\pm$    0.68 &   3.587 $\pm$   0.063 &    0.969 $\pm$   0.124 &	0.1  &   1.298 &   13.0  &    3.27 $\pm$	0.57 \\ 
\hline
B1B2-MM1 &    3.03 $\pm$    0.30 &   3.633 $\pm$   0.076 &    1.697 $\pm$   0.199 &	0.1  &   0.476 &   13.3  &    3.04 $\pm$	0.47 \\ 
B1B2-MM2 &    3.34 $\pm$    0.26 &   4.148 $\pm$   0.063 &    1.931 $\pm$   0.190 &	0.1  &   0.419 &   15.8  &    3.69 $\pm$	0.46 \\ 
\hline
B2-MM1	 &    3.62 $\pm$    0.39 &   3.826 $\pm$   0.053 &    1.041 $\pm$   0.125 &	0.1  &   0.430 &   14.1  &    2.20 $\pm$	0.35 \\ 
B2-MM2   &    1.88 $\pm$    0.14 &   3.782 $\pm$   0.047 &    1.072 $\pm$   0.086 &     0.1 &   0.301 &   14.2  &    1.18 $\pm$    0.13 \\ 
B2-MM8	 &    1.70 $\pm$    0.21 &   3.380 $\pm$   0.121 &    1.875 $\pm$   0.248 &	0.1  &   0.416 &   14.5  &    1.84 $\pm$	0.33 \\ 
B2-MM14  &    1.90 $\pm$    0.30 &   3.676 $\pm$   0.137 &    1.632 $\pm$   0.278 &	0.1  &   0.519 &   14.6  &    1.80 $\pm$	0.42 \\ 
B2-MM16  &    2.73 $\pm$    0.14 &   3.662 $\pm$   0.032 &    1.162 $\pm$   0.065 &	0.1  &   0.181 &   14.3  &    1.85 $\pm$	0.14 \\ 
\hline
C-W	 &    3.70 $\pm$    0.33 &   3.698 $\pm$   0.329 &    1.626 $\pm$   0.167 &	0.1  &   0.531 &   12.0  &    3.71 $\pm$	0.50 \\ 
C-MM3	 &    1.97 $\pm$    0.24 &   3.906 $\pm$   0.070 &    2.602 $\pm$   0.231 &	0.1  &   0.259 &   12.3  &    3.12 $\pm$	0.47 \\ 
	 &    5.31 $\pm$    0.36 &   3.657 $\pm$   0.014 &    0.500 $\pm$   0.042 &	0.1  &   0.259 &   12.3  &    1.62 $\pm$	0.18 \\ 
C-MM4	 &    8.91 $\pm$    0.77 &   3.716 $\pm$   0.030 &    0.765 $\pm$   0.081 &	0.1  &   0.662 &   10.4  &    4.57 $\pm$	0.62 \\ 
C-MM5	 &    5.31 $\pm$    0.36 &   3.657 $\pm$   0.014 &    0.499 $\pm$   0.042 &	0.1  &   0.241 &   10.7  &    1.74 $\pm$	0.19 \\ 
	 &    1.97 $\pm$    0.24 &   3.905 $\pm$   0.070 &    2.599 $\pm$   0.231 &	0.1  &   0.241 &   10.7  &    3.37 $\pm$	0.50 \\ 
C-MM6	 &    3.99 $\pm$    0.62 &   3.605 $\pm$   0.035 &    0.557 $\pm$   0.090 &	0.1  &   0.411 &   13.0  &    1.33 $\pm$	0.30 \\ 
	 &    2.52 $\pm$    0.40 &   3.702 $\pm$   0.090 &    2.423 $\pm$   0.242 &	0.1  &   0.411 &   13.0  &    3.64 $\pm$	0.68 \\ 
C-MM7	 &    2.97 $\pm$    0.36 &   3.772 $\pm$   0.105 &    1.593 $\pm$   0.208 &	0.1  &   0.607 &   12.0  &    2.92 $\pm$	0.52 \\ 
\hline
E-MM1	 &    4.51 $\pm$    0.61 &   3.764 $\pm$   0.053 &    0.706 $\pm$   0.165 &	0.1  &   0.505 &   15.0  &    1.83 $\pm$	0.50 \\ 
	 &    2.57 $\pm$    0.54 &   4.694 $\pm$   0.106 &    0.681 $\pm$   0.199 &	0.1  &   0.505 &   15.0  &    1.01 $\pm$	0.36 \\ 
E-MM2d	 &    3.10 $\pm$    0.27 &   3.823 $\pm$   0.047 &    1.324 $\pm$   0.136 &	0.1  &   0.322 &   13.6  &    2.41 $\pm$	0.32 \\ 
E-MM3	 &    2.94 $\pm$    0.27 &   3.449 $\pm$   0.099 &    2.580 $\pm$   0.308 &	0.1  &   0.907 &   15.0  &    4.37 $\pm$	0.66 \\ 
E-MM4	 &    3.26 $\pm$    0.20 &   3.604 $\pm$   0.049 &    1.708 $\pm$   0.128 &	0.1  &   0.301 &   15.0  &    3.21 $\pm$	0.31 \\ 
E-MM5	 &    3.90 $\pm$    0.30 &   3.882 $\pm$   0.068 &    2.030 $\pm$   0.191 &	0.1  &   0.859 &   15.0  &    4.56 $\pm$	0.55 \\ 
\hline
F-MM1	 &    4.53 $\pm$    0.19 &   3.717 $\pm$   0.028 &    1.324 $\pm$   0.070 &	0.1  &   0.252 &   15.3  &    3.44 $\pm$	0.23 \\ 
	 &    3.20 $\pm$    0.44 &   4.732 $\pm$   0.017 &    0.222 $\pm$   0.021 &	0.1  &   0.252 &   15.3  &    0.41 $\pm$	0.07 \\ 
F-MM2	 &    4.10 $\pm$    0.18 &   3.774 $\pm$   0.030 &    1.517 $\pm$   0.079 &	0.1  &   0.233 &   15.6  &    3.56 $\pm$	0.24 \\ 
\hline
\end{tabular}
\end{table*}

\begin{table*}
{\caption{Column density, deuterium fraction and CO-depletion factor}\label{N-RD-fD-results}
{\small
\begin{tabular}{lccccccccc}

\hline
\hline
\\
Source & N$_{tot}$(N$_2$H$^+$) & N$_{tot}$(N$_2$D$^+$) & R$_{\rm D}$ & S$_{850\mu m}$ &T$_k^a$ &T$_k$& N(H$_2$)&  N$_{tot}$(C$^{17}$O) & f$_{\rm d}$ \\
       & (10$^{13}$~cm$^{-2}$) & (10$^{12}$~cm$^{-2}$) &       &(Jy beam$^{-1}$)& (K)    & ref.$^b$  & (10$^{22}$~cm$^{-2}$) & (10$^{15}$~cm$^{-2}$)  &\\
\hline
\hline
A3-MM1	&0.22	 $\pm$ 0.02 	&	<0.20		&	<0.09	&  262 $\pm$    35 &    17.6 $\pm$ 0.7 & P & 1.00 $\pm$ 0.13 &   1.32 $\pm$    0.24 		   &   0.66 $\pm$   0.15 \\ 
A-MM4	&1.19	 $\pm$ 0.28 	&	<0.26		&	<0.02	&  541 $\pm$    46 &    16.3 $\pm$ 0.5 & P & 2.34 $\pm$ 0.20 &   1.92 $\pm$    0.20 		   &   1.06 $\pm$   0.14 \\ 
A-MM5	&1.10	 $\pm$ 0.06 	&	<0.23		&	<0.02	& 1035 $\pm$    59 &    18.6 $\pm$ 0.7 & P & 3.64 $\pm$ 0.21 &   2.37 $\pm$    0.30 		   &   1.33 $\pm$   0.18 \\ 
VLA1623	&1.60	 $\pm$ 0.19 &1.57	 $\pm$ 0.13 &0.10	 $\pm$ 0.01 & 4887 $\pm$   108 &    16.4 $\pm$ 0.5 & P & 20.89 $\pm$ 0.46 &   3.13 $\pm$    0.13 		   &   5.80 $\pm$   0.27 \\ 
SM1N	&4.21	 $\pm$ 0.37 &1.59	 $\pm$ 0.08 &0.038	 $\pm$ 0.004& 5760 $\pm$    77 &    17.3 $\pm$ 0.6 & P & 22.63 $\pm$ 0.30 &   6.20 $\pm$    0.31 		   &   3.17 $\pm$   0.16 \\ 
SM1  	&3.82	 $\pm$ 0.30 &2.56	 $\pm$ 0.14 &0.067	 $\pm$ 0.006& 8601 $\pm$    84 &    17.2 $\pm$ 0.6 & P & 34.10 $\pm$ 0.33 &   6.86 $\pm$    0.13 		   &   4.32 $\pm$   0.09 \\ 
A-MM6	&0.51	 $\pm$ 0.09 	&	<0.26		&	<0.05	& 1864 $\pm$    67 &    18.8  $\pm$ 0.8 & P & 6.44 $\pm$ 0.23 &   4.24 $\pm$    0.23 		   &   1.32 $\pm$   0.09 \\ 
SM2	&1.65	 $\pm$ 0.15 &0.97	 $\pm$ 0.10 &0.059	 $\pm$ 0.008& 3746 $\pm$    45 &    18.5  $\pm$ 0.7 & P & 13.27 $\pm$ 0.16 &   4.28 $\pm$    0.20 	   &   2.69 $\pm$   0.13 \\ 
A-MM8	&1.19	 $\pm$ 0.12 	&0.29    $\pm$ 0.08		&0.024   $\pm$ 0.007	&	1065 $\pm$ 42	       &    18.4 $\pm$ 0.7 & P & 3.80 $\pm$ 0.15 &\\  
A-S		&0.11	 $\pm$ 0.03 	&	<0.25		&	<0.23		&	109 $\pm$ 40	       &    20 & M & 0.34 $\pm$ 0.13 & \\  
\hline											
B1-MM1	&0.24	 $\pm$ 0.06 	&0.93	 $\pm$ 0.09 	&0.39	 $\pm$ 0.10 	        &  131 $\pm$    39 &    16.3 $\pm$ 1.9 & F & 0.57 $\pm$ 0.17 &    2.29 $\pm$    0.12 &   0.22 $\pm$   0.07 \\ 
B1-MM3	&1.92	 $\pm$ 0.31 	&6.22	 $\pm$ 1.62		&0.32	$\pm$	0.10	&  706 $\pm$    20 &    12.2 $\pm$ 0.3 & P & 5.02 $\pm$ 0.14 &   2.25 $\pm$    0.11 &   1.94 $\pm$   0.11 \\ 
B1-MM4	&3.33	 $\pm$ 0.47 	&4.15	 $\pm$ 0.27 	&0.12	 $\pm$ 0.02 	        &  528 $\pm$    31 &    11.8 $\pm$ 0.2 & P & 3.99 $\pm$ 0.23 &   2.87 $\pm$    0.24 &   1.21 $\pm$   0.12 \\ 
\hline											
B1B2-MM1&0.51	 $\pm$ 0.03 	&1.24	 $\pm$ 0.08 	&0.24	 $\pm$ 0.02 	        &  216 $\pm$    47 &    13.3 $\pm$ 1.2 & F & 1.31 $\pm$ 0.28 &   2.09 $\pm$    0.12 &   0.55 $\pm$   0.12 \\ 
B1B2-MM2&0.30	 $\pm$ 0.09 	&	<0.26			&	<0.09		&  107 $\pm$    27 &    15.8 $\pm$ 0.5 & P &  0.49 $\pm$ 0.12 &  3.08 $\pm$    0.13 &   0.14 $\pm$   0.03 \\ 
\hline											
B2-MM1	&0.87	 $\pm$ 0.24 	&1.22	 $\pm$ 0.09 	&0.14	 $\pm$ 0.04 	        &  207 $\pm$    39 &    14.1 & F & 1.14 $\pm$ 0.21 &   1.46 $\pm$    0.15 	   &   0.68 $\pm$   0.15 \\ 
B2-MM2	&1.21	 $\pm$ 0.25 	&1.46	 $\pm$ 0.15 	&0.12	 $\pm$ 0.03 	        &  470 $\pm$    40 &    11.4 $\pm$ 0.2 & P & 3.80 $\pm$ 0.32 &   1.08 $\pm$    0.12 &   3.06 $\pm$   0.43 \\ 
B2-MM6	&1.88	 $\pm$ 0.43 	&3.34	 $\pm$ 0.15 	&0.18	 $\pm$ 0.04 	        & 1190 $\pm$    35 &    11.3 $\pm$ 0.2 & P & 9.78 $\pm$ 0.28 &   1.23 $\pm$    0.13 &   6.91 $\pm$   0.76 \\ 
B2-MM8	&2.70	 $\pm$ 0.28 	&1.85	 $\pm$ 0.23 	&0.07	 $\pm$ 0.01 	        & 1447 $\pm$    34 &    13.5 $\pm$ 0.4 & P & 8.57 $\pm$ 0.20 &   1.87 $\pm$    0.32 &   3.98 $\pm$   0.69 \\ 
B2-MM10	&0.95	 $\pm$ 0.09 	&	<0.23		&	<0.02		        & 1026 $\pm$    41 &    15.8 $\pm$ 0.5 & P & 4.66 $\pm$ 0.19 &   2.02 $\pm$    0.15 &   2.00 $\pm$   0.17 \\ 
B2-MM14	&1.82	 $\pm$ 0.53 	&1.67	 $\pm$ 0.17 	&0.09	 $\pm$ 0.03 	        & 1135 $\pm$    29 &    10.7 $\pm$ 0.2 & P & 10.38 $\pm$ 0.27 &   1.35 $\pm$    0.21 &   6.68 $\pm$   1.05 \\ 
B2-MM15	&1.80	 $\pm$ 0.38 	&0.61	 $\pm$ 0.10 	&0.03	 $\pm$ 0.01 	        &  834 $\pm$    32 &    11.8 $\pm$ 0.3 & P & 6.31 $\pm$ 0.24 &   1.33 $\pm$    0.20 &   4.12 $\pm$   0.64 \\ 
B2-MM16	&1.54	 $\pm$ 0.37 	&2.11	 $\pm$ 0.11 	&0.14	 $\pm$ 0.03 	        & 1001 $\pm$    30 &    10.4 $\pm$ 0.2 & P & 9.69 $\pm$ 0.29 &   1.36 $\pm$    0.08 &   6.19 $\pm$   0.41 \\ 
B2-MM17	&1.36	 $\pm$ 0.17 	&	---			&	---		&  864 $\pm$    30 &    13.4 $\pm$ 0.3 & P & 5.19 $\pm$ 0.18 &   2.00 $\pm$    0.26 &   2.25 $\pm$   0.30 \\ 
\hline											
C-We	&0.48	 $\pm$ 0.13 	&0.75	 $\pm$ 0.07 	&0.16	 $\pm$ 0.04 	        &  179 $\pm$    34 &    12 & M & 1.31 $\pm$ 0.25 &   3.38 $\pm$    0.09             &   0.34 $\pm$   0.07 \\ 
C-Ne	&1.23	 $\pm$ 0.15 	&5.25	 $\pm$ 1.08 	&0.43	 $\pm$ 0.10 	        &  412 $\pm$    27 &    12 & M & 3.02 $\pm$ 0.20 &   3.21 $\pm$    0.15             &   0.82 $\pm$   0.07 \\ 
C-MM3	&1.67	 $\pm$ 0.19 	&	---			&	---		& 1224 $\pm$    29 &    12.3 $\pm$ 0.3 & P & 8.57 $\pm$ 0.20 &   4.32 $\pm$    0.23 &   1.72 $\pm$   0.10 \\ 
C-MM4	&1.84	 $\pm$ 0.56 	&	---			&	---		& 1066 $\pm$    32 &    10.4 $\pm$ 0.7 & F & 10.32 $\pm$ 0.31 &   4.48 $\pm$    0.25 &   2.00 $\pm$   0.13 \\ 
C-MM5	&2.29	 $\pm$ 0.29 	&3.93	 $\pm$ 1.29 	&0.17	 $\pm$ 0.06 	        & 1176 $\pm$    29 &    10.7 $\pm$ 0.8 & F & 10.75 $\pm$ 0.27 &   4.59 $\pm$    0.49 &   2.04 $\pm$   0.22 \\ 
C-MM6	&2.10	 $\pm$ 0.28 	&	---			&	---		&  976 $\pm$    31 &    12.8 $\pm$ 0.4 & P & 6.18 $\pm$ 0.20 &   4.55 $\pm$    0.22 &   1.18 $\pm$   0.07 \\ 
C-MM7	&2.12	 $\pm$ 0.33 	&	---			&	---		&  450 $\pm$    42 &    12 & M & 3.30 $\pm$ 0.30 &   3.77 $\pm$    0.27             &   0.76 $\pm$   0.09 \\ 
\hline											
E-MM1e	&0.12	 $\pm$ 0.01 	&	<0.22		&	<0.18		        &	34 $\pm$ 36	       &    15 & M & 0.17 $\pm$ 0.18 &\\  
E-MM2d	&0.59	 $\pm$ 0.17 	&4.51	 $\pm$ 1.30 	&0.42	 $\pm$ 0.02 	        &  655 $\pm$    36 &    13.6 $\pm$ 0.4 & P & 3.83 $\pm$ 0.21 &   3.05 $\pm$    0.07 &   1.09 $\pm$   0.07 \\ 
E-MM3	&	---			&	---			&	---	&  217 $\pm$    31 &    15 & M & 1.07 $\pm$ 0.15 &   4.19 $\pm$    0.50             &   0.22 $\pm$   0.04 \\ 
E-MM4	&0.29	 $\pm$ 0.09 	&0.55	 $\pm$ 0.07 	&0.19	 $\pm$ 0.06 	        &  369 $\pm$    53 &    15 & M & 1.83 $\pm$ 0.26 &   3.60 $\pm$    0.13             &   0.44 $\pm$   0.07 \\ 
E-MM5	&	---			&	---			&	---	&  363 $\pm$    41 &    15 & M & 1.79 $\pm$ 0.20 &   3.88 $\pm$    0.30             &   0.40 $\pm$   0.05 \\ 
\hline											
F-MM1	&0.87	 $\pm$ 0.14 	&	<0.18		&	<0.02		        &  741 $\pm$    41 &    15.3 $\pm$ 0.5 & P & 3.55 $\pm$ 0.20 &   4.27 $\pm$    0.23 &   0.72 $\pm$   0.06 \\ 
F-MM2	&1.18	 $\pm$ 0.15 	&1.54	 $\pm$ 0.19 	&0.13	 $\pm$ 0.02 	        &  760 $\pm$    41 &    15.6 $\pm$ 0.5 & P & 3.52 $\pm$ 0.19 &   3.68 $\pm$    0.22 &   0.83 $\pm$   0.07 \\ 
\hline											
H-MM1	&0.88	 $\pm$ 0.09 	&3.76	 $\pm$ 0.86 	&0.43	 $\pm$ 0.11 	        &  614 $\pm$    99 &    11 $\pm$ 0.2 & P & 5.32 $\pm$ 0.86 &   1.04 $\pm$    0.15   &   4.44 $\pm$   0.96 \\ 
I-MM1	&1.09	 $\pm$ 0.13 	&3.40	 $\pm$ 0.89 	&0.31	 $\pm$ 0.09 	        &  392 $\pm$   135 &    11 & & 4.11 $\pm$ 1.41 &   0.49 $\pm$    0.06               &   7.30 $\pm$   2.66 \\ 
\hline

\end{tabular}\\
}
$^a$We assume gas and dust to have the same temperature. $^b$Temperature taken from: P --- \citet{Pattle2015}, F --- \citet{Friesen2009}, M --- \citet{Motte1998}.
}
\end{table*}